\documentclass[prc,aps,floats,floatfix,showpacs,nofootinbib,%
superscriptaddress]{revtex4}

\usepackage{amsmath}
\usepackage{amssymb}
\usepackage{amsfonts}
\usepackage{graphicx}
\usepackage{tabularx}
\usepackage{multirow}
\usepackage{lscape}
\usepackage{rotating}
\usepackage{pstricks}

\begin{document}

\title{Linear response theory in asymmetric nuclear matter for Skyrme functionals including spin-orbit and tensor terms.}


\author{D. Davesne}
\email{davesne@ipnl.in2p3.fr}
\affiliation{Universit\'e de Lyon, F-69003 Lyon, France; Universit\'e Lyon 1,
             43 Bd. du 11 Novembre 1918, F-69622 Villeurbanne cedex, France\\
             CNRS-IN2P3, UMR 5822, Institut de Physique Nucl{\'e}aire de Lyon}
\author{A. Pastore}
\affiliation{Institut d'Astronomie et d'Astrophysique, CP 226, Universit\'e Libre de Bruxelles, B-1050 Bruxelles, Belgium}
\author{J. Navarro}
\affiliation{IFIC (CSIC-Universidad de Valencia), Apartado Postal 22085, E-46.071-Valencia, Spain}


\begin{abstract}
The formalism of linear response theory for a Skyrme functional including spin-orbit and tensor terms is generalized to the case of infinite nuclear matter with arbitrary isospin asymmetry. Response functions are obtained by solving an algebraic system of equations, which is explicitly given. Spin-isospin strength functions are analyzed varying the conditions of density, momentum transfer, asymmetry and temperature. The presence of instabilities, including the spinodal one, is studied by means of the static susceptibility. 
\end{abstract}


 
\date{\today}


\maketitle

\section{Introduction}
\label{sect:intro}

The energy density functional (EDF) is a tool of choice for a systematic and quantitative description of properties of atomic nuclei from drip-line to drip-line~\cite{ben03}. Among the non-relativistic EDF those related to the zero-range non-local Skyrme interaction are the most widely used. In its standard form~\cite{vau72}, Skyrme's pseudo-potential contains central, spin-orbit and density-dependent terms. Tensor terms, present in Skyrme's initial proposal~\cite{sky59}, have been included only recently~\cite{les07,kor13,sag14}, and show to play a major role in finite nuclei, both for ground state properties~\cite{les07} and excited states~\cite{cao11}.  The resulting EDF~\cite{per04} is written as a linear combination of local densities, whose coupling constants are optimally determined by minimizing a multi-dimensional merit function for a given set of observables or pseudo-observables~\cite{kor10}. This procedure is not simple and requires an accurate selection of the observables included in the fit to guarantee a proper constraint for each coupling constant. 

The response of atomic nuclei to different probes is the most efficient way to obtain information about the structure and specific manifestations of the effective nucleon-nucleon interaction in the nuclear medium. 
Physical insight can be obtained from INM which, as a homogeneous medium made of interacting nucleons, is an ideal but very useful system, largely employed because of its relative simplicity. 
This model  is not only connected with the inner part of atomic nuclei, but it is also very useful to describe some phenomena in the interior of compact stars~\cite{gal11,cha13,gul13}.
Most of the investigations of responses have been based on random phase approximation (RPA) or linear response (LR) theory~\cite{fet71,rin80}, in which excitations result from the residual particle-hole (ph) interaction between particles below and above the Fermi level. 

Working with a Skyrme functional in infinite nuclear matter (INM) allows one to derive analytic expressions of the nuclear response function, thus making numerical calculations much more rapid as compared to those for finite nuclei. In this respect, several groups have investigated the properties of infinite systems trying to find additional constraints for the coupling constants of the functional. A typical example is the use of Landau inequalities to avoid the appearance of instabilities in the infinite system at low momentum~\cite{mar02,cao10,nav13}. 
The Landau-Migdal approach is only valid in the so-called \emph{long-wavelength limit}, thus it is not able to detect eventual \emph{finite-size} instabilities that occur at nonzero values of transferred momentum 
$q$~\cite{les06,kor10b,hel13}.
The additional constraint coming from the calculations of the response function in INM could be thus used to avoid such kind of problems and further reduce the parameter space one need to explore during the optimization procedure~\cite{kor13,kor10,pas12c}.

The response of atomic nuclei to different probes is the most efficient way to obtain information about the structure and specific manifestations of the effective nucleon-nucleon interaction in the nuclear medium. 
Physical insight can be obtained from INM which, as a homogeneous medium made of interacting nucleons, is an ideal but very useful system, largely employed because of its relative simplicity. 
This model  is not only connected with the inner part of atomic nuclei, but it is also very useful to describe some phenomena in the interior of compact stars~\cite{gal11,cha13,gul13}.
Most of the investigations of responses have been based on random phase approximation (RPA) or linear response (LR) theory~\cite{fet71,rin80}, in which excitations result from the residual particle-hole (ph) interaction between particles below and above the Fermi level. 

In the present article, the EDF described in \cite{per04} will be our starting point and it will be used for the determination of spin-isospin response functions. We have decided to use the Skyrme EDF instead of the interaction form because in this way the formalism can be easily generalized to recent extensions of the functional~\cite{cha09,mar12,car08,sad13,dav13}. 
The RPA response of symmetric nuclear matter (SNM) has been presented for the case of the standard Skyrme pseudo-potential in~\cite{gar92}, and extended to include spin-orbit~\cite{mar06} and tensor terms~\cite{dav09}. The formalism has been revised in~\cite{pas12,pas12b} to treat the case of a more general Skyrme energy density functional (EDF)~\cite{per04}, both for SNM and pure neutron matter (PNM). 

Since neutron excess is a common situation in finite nuclei and also of current interest in astrophysical context, a generalization to asymmetric nuclear matter (ANM) is clearly needed. The RPA response for such a system  has been derived in~\cite{her97,her97a} considering only the central part of a Skyrme interaction. The purpose of this work is to extend the formalism by including spin-orbit and tensor terms. We will show that, analogously to the SNM case, the specific momentum dependence of the Skyrme EDF leads to an algebraic system of coupled equations, with however a double number of equations, from which one can deduce the RPA response function for the required spin and isospin channel. We present results for different isospin asymmetry, density and transferred momentum. Thermal effects on response functions are also discussed. However, we restrict ourselves to isospin excitations induced by the operator $\tau_3$, the case of charge exchange processes being left for a future study.  

The article is organized as follows. In Sec.~\ref{matrix:el} we give the matrix elements in the spin-isospin space of the residual interaction between particles and holes (ph) obtained from a general Skyrme EDF. The basic formalism to obtain the RPA response function is presented in Sec.~\ref{sec:form}. In Sec.~\ref{sec:res} we give and discuss the main results, including the presence of instabilities. In Sec.~\ref{sec:conclusion} we draw our conclusions. Finally, the algebraic system of equations and some technical details are given in the Appendices.

\section{The residual particle-hole interaction}\label{matrix:el}

In this section, we give explicit expressions for the matrix elements of the residual particle-hole (ph) interaction, $V_{ph}$. Within the context of EDF and LR theory \cite{rin80},  they can be obtained by performing a  second functional derivative of the total energy as
\begin{equation}
\langle ({\mathbf q}_1, \sigma_1, \tau_1), ({\mathbf q}_3, \sigma_3, \tau_3)^{-1} | V_{ph} | 
({\mathbf q}_4, \sigma_4, \tau_4), ({\mathbf q}_2, \sigma_2, \tau_2)^{-1} \rangle 
= \frac{\delta^2 E}{\delta \rho_{31} \delta \rho_{42}} \, ,
\end{equation}
where $({\mathbf q}, \sigma, \tau)$ refer to the momentum and spin and isospin projections of the corresponding particle or hole, and  $\rho_{ij}$ stands for the occupation number matrix.  Due to momentum conservation, there are only three independent momenta, which we choose as those of the holes ${\bf k}_1={\bf q}_3$ and ${\bf k}_2={\bf q}_2$, and the transferred momentum ${\bf q}$, using the standard notation \cite{gar92}. Each p-h pair must be coupled to well-defined values of spin $S$ and projection $M$.
 Regarding the isospin, it is more convenient in ANM to work in the proton/neutron formalism. Dropping  momentarily the momenta and spin dependence, we thus need the matrix elements 
$\langle \tau_1, \tau_3^{-1} | V_{ph} | \tau_4, \tau_2^{-1} \rangle$, where the indices $\tau_i$ refer to protons ($p$) or neutrons ($n$).
 Due to charge conservation, there are six matrix elements, namely 
$$\langle p, p^{-1} | V_{ph} | p, p^{-1} \rangle, \langle n, n^{-1} | V_{ph} | n, n^{-1} \rangle, 
\langle p, p^{-1} | V_{ph} | n, n^{-1} \rangle, \langle n, n^{-1} | V_{ph} | p, p^{-1} \rangle,$$
$$ \langle p, n^{-1} | V_{ph} | n, p^{-1} \rangle, 
\langle n, p^{-1} | V_{ph} | p, n^{-1} \rangle.$$
The latter two matrix elements are only relevant for situations involving charge exchange reactions. They have been already  discussed in \cite{her99}, and we will not consider them in the present article.
 The former four matrix elements will be written as 
$\langle \tau, \tau^{-1} | V_{ph} | \tau', {\tau'}^{-1} \rangle$. 
Actually, as $\langle p, p^{-1} | V_{ph} | n, n^{-1} \rangle = \langle n, n^{-1} | V_{ph} | p, p^{-1} \rangle$, there are only three independent elements. To avoid repetition of indices we shall write them as
\begin{equation}
V_{ph}^{(\tau SM;\tau' S'M')}({\mathbf k}_1, {\mathbf k}_2)
= \langle \tau, \tau, SM | V_{ph}({\mathbf k}_1, {\mathbf k}_2)| \tau',\tau' S'M'\rangle\,.
\end{equation}

From now on, we specialize to the Skyrme functional defined in \cite{per04}, which includes both spin-orbit and tensor terms. We can write 
\begin{eqnarray}\label{mel:nn}
V_{ph}^{(\tau SM;\tau' S'M')}({\mathbf k}_1, {\mathbf k}_2)&=&
\frac{1}{2}  \delta_{SS'}\delta_{MM'} 
\left( W_1^{(\tau,\tau',S)} + W_2^{(\tau,\tau',S)} {\mathbf k}_{12}^2  \right) 
\nonumber \\
&+& \frac{1}{2}  \delta_{SS'} \delta_{S1}  W_{T1}^{(\tau,\tau')} \, (-)^M (k_{12})^{(1)}_{-M} (k_{12})^{(1)}_{M'}  \nonumber \\
&+& \frac{1}{2} W_{SO}^{(\tau,\tau')} 
\left( \delta_{S'0}\delta_{S1}M(k_{12})^{(1)}_{-M}+\delta_{S'1}\delta_{S0}M'(k_{12})^{(1)}_{M'}\right)\, ,
\end{eqnarray}
where ${\mathbf k}_{12}$ is the relative hole momentum, and following the notation of~\cite{dav09}, we have introduced the rank-1 tensor 
\begin{equation}
(k_{12})^{(1)}_{\mu}=\sqrt{\frac{4\pi}{3}}[k_{1\mu}Y_{1\mu}(\hat{k}_1)-k_{2\mu}Y_{1\mu}(\hat{k}_2)]\,.
\end{equation}
The coefficients $W_{i}^{(\tau,\tau',S)}$ are combinations of the coupling constants of the functional, and are given in appendix \ref{app:w}. Some of them depend on the transferred momentum ${\mathbf q}$ and also on the isoscalar and isovector densities, $\rho_0$ and $\rho_1$ respectively. From Eq. (\ref{mel:nn}) we observe that, analogously to the symmetric matter case~\cite{dav09}, the tensor term (second line) acts only in the $S=1$ channel, and couples different projections $M$ and $M'$, while the spin-orbit term (last line) couples both spin channels. 


\section{Response functions}\label{sec:form}
The method for calculating RPA response functions has been presented in Refs. \cite{gar92,mar06,dav09} for SNM. It consists in first solving the Bethe-Salpeter equation for the RPA p-h propagator, taking advantage of the particular form of the p-h interaction in momentum space, and afterwards averaging the p-h propagator over momenta to get the response function. Some details on the method have recently been given in \cite{pas14}. Here we generalize it for ANM, which will be characterized by the asymmetry parameter $Y=\rho_1/\rho_0$, defined as the quotient between the isoscalar and isovector densities.

The first required ingredient is the Hartree-Fock (HF) retarded propagator of a non-interacting p-h pair. As charge-exchange processes are not considered here, the particle and the hole in the same pair share the same isospin number $\tau$ (either $p$ or $n$). The HF p-h propagator can thus be written as
\begin{equation}
G^{(\tau)}_{HF}(\mathbf{k},\mathbf{q},\omega) = \frac{n_{\tau}(\mathbf{k})-n_{\tau}(\mathbf{q}+\mathbf{k})}{ \omega +\varepsilon_{\tau}(\mathbf{k})- \varepsilon_{\tau}(\mathbf{q}+\mathbf{k})+i \eta} \, ,
 \label{GHF}
\end{equation} 
where $n_{\tau}(\mathbf{k})$ is the Fermi-Dirac occupation number of $\tau$-particles, given by the step function $\theta(k^{\tau}_F-k)$ at zero temperature, and  $\varepsilon_{\tau}(\mathbf{k})$ is the single-particle energy
\begin{equation}
\varepsilon_{\tau}(\mathbf{k}) = \frac{k^2}{2 m_{\tau}^*} + U_{\tau}
\end{equation} 
where $U_{\tau}$ is the mean field, excluding the $k^2$ dependence contributing to the effective mass $m_{\tau}^*$, written as
\begin{eqnarray}
\frac{1}{2m_n^*} & = & \frac{1}{2m_n} + C^{\tau}_0 \rho + Y C^{\tau}_1 \rho \, , \\
\frac{1}{2m_p^*} & = & \frac{1}{2m_p} + C^{\tau}_0 \rho - Y C^{\tau}_1 \rho \, ,
\end{eqnarray}
where $C^{\tau}_{0,1}$ are EDF coupling constants. Notice the use of natural units ($\hbar=c=1$), as will be done along this article. From now on, vector ${\mathbf q}$ will define the $z$-axis. 

The HF response function is obtained as
\begin{equation} 
\chi^{(\tau)}_{HF}(q,\omega) = 2 \int \frac{d^3 \mathbf{k}}{(2 \pi)^3} \, G^{(\tau)}_{HF}(\mathbf{k},q,\omega) \, ,
\label{chiHF}
\end{equation} 
where the factor 2 stands for the spin-degeneracy. In the following we will often deal with momentum averages similar to the previous one, which will be indicated within brackets: $\chi^{(\tau)}_{HF}(q,\omega) = 2 
 \langle G^{(\tau)}_{HF} \rangle$. 
 
Notice that the HF propagator is independent of the spin-channel, and so is the response $\chi^{(\tau)}_{HF}(q,\omega)$. The correlated RPA p-h propagator depends  however on the p-h spin and isospin quantum numbers. Since we do not consider charge-exchange processes we can write  the Bethe-Salpeter equation as
\begin{eqnarray}
G^{(\tau \tau' SM)}(\mathbf{k}_1,q,\omega) &=& 
\delta(\tau,\tau') G_{HF}^{(\tau)}(\mathbf{k}_1,q,\omega) \nonumber \\
&+& 
G^{(\tau)}_{HF}(\mathbf{k}_1,q,\omega) 
\sum_{(\tau'',S''M'')} \int \frac{d^{3}\mathbf{k}_2}{(2 \pi)^3} \,
V_{ph}^{(\tau S M;\tau'' S''M'')}({\mathbf k}_1, {\mathbf k}_2)
G^{(\tau'' \tau' S''M'')}(\mathbf{k}_2,q,\omega) \, , 
\label{BSeq}
\end{eqnarray}
The residual interaction links two p-h pairs with quantum numbers $(\tau SM)$ and $(\tau' S'M')$, and hole momenta $\mathbf{k}_1$ and $\mathbf{k}_2$, respectively. We refer to  Ref. \cite{her97a} for more details about the adopted notation.  The linear response function is obtained as
\begin{equation}
\chi^{(\tau \tau' SM)}(q,\omega)  = 2  \langle G^{(\tau \tau' SM)} \rangle \, .
\end{equation}

By inspecting the p-h interaction (\ref{mel:nn}), one can see that a closed system of algebraic equations is obtained by multiplying Eq. (\ref{BSeq}) successively with the functions $1$, $k^2$, $k Y_{1,0}$, $k^2 |Y_{1,\pm 1}| ^2$, and $k^2 |Y_{1,0}| ^2$,  and integrating over the momentum $\mathbf{k}_1$. As compared to the SNM case, the number of equations is doubled because of the isospin indices $\tau,\tau'$.
The tensor term $W_{T1}^{(\tau \tau')}$ is only effective on the $S=1$ channel, but it can also influence the $S=0$ channel, due to the mixing between both spin channels induced by the spin-orbit term $W_{SO}$. 
Similarly to SNM \cite{mar06,dav09} this coupling can be absorbed into an effective coefficient $\widetilde{W}_1^{(\tau \tau',S)}$ (cf. Appendix \ref{app-wp}), so that we deal in practice with two separate systems for each spin channel. 

For fixed values of the spin quantum numbers $(S,M)$ there are four possible isospin combinations, namely $(nn), (pn), (np), (pp)$. However, due to isospin properties of the residual interaction, equations can be actually decoupled in two subsystems. One of them is for the couple $(nn)-(pn)$ and it can be written in matrix form as
\begin{eqnarray}
\left( \begin{matrix}
A_{nn} & A_{np} \\
A_{pn} & A_{pp} \\
\end{matrix} \right) \,
\left( \begin{matrix}
X_{nn}  \\
X_{pn}  \\
\end{matrix} \right)
= \left( \begin{matrix}
B_n  \\
 0 \\
\end{matrix} \right)\,,
\label{pepet}
\end{eqnarray}
where the column vector $X_{\tau \tau'}$ contains the unkown momentum averages of the RPA propagator, 
$A_{\tau \tau'}$ are square matrices which depend on the EDF coupling constants and averages of the HF propagator, and $B_{\tau}$ are column vectors depending on HF averages. The explicit expressions of these quantities are given in Appendix \ref{app:EQS}. The other subsystem is for the couple $(pp)-(np)$, which is obtained from the previous one by simply replacing $n \leftrightarrow p$.  The number of coupled equations for each subsystem is 6 for the channel $S=0$, and 8 for $S=1$. The expressions become cumbersome, preventing us to write the response function in a compact form as in SNM~\cite{dav09} or PNM~\cite{pas12b}. Instead, it is preferable to numerically  solve these systems in the $(q,\omega)$-space. We have nevertheless relied on the analytical systems to derive some interesting quantities, as for instance sum rules.

As discussed in~\cite{her97,her97a}, the relevant  spin-isospin responses are given by the combinations
\begin{eqnarray}
\chi^{(SM;I=0)}(q, \omega) &=& 
\chi^{(nnSM)}(q, \omega) + \chi^{(pnSM)}(q, \omega) 
+ \chi^{(ppSM)}(q, \omega) + \chi^{(npSM)}(q, \omega)\,, \\
\chi^{(SM;I=1)}(q, \omega) &=& 
\chi^{(nnSM)}(q, \omega) - \chi^{(pnSM)}(q, \omega) 
+ \chi^{(ppSM)}(q, \omega) - \chi^{(npSM)}(q, \omega) \,.
\end{eqnarray}
Actually, instead of the response functions we deal with the corresponding strength functions, defined as
\begin{equation}
S^{(S,M,I)}(q,\omega) = -\frac{1}{\pi} {\rm Im} \chi^{(S,M,I)} \, ,
\end{equation}
since all physical properties are actually embedded into it.

\section{Results}\label{sec:res}

To present our results, we have chosen the asymmetry parameter values $Y=0.21$, and 0.5, which roughly correspond to the  isospin asymmetry of $^{208}$Pb and the $\beta$-equilibrium condition, respectively. To complete the discussion we will also show the extreme cases of SNM ($Y=0$) and PNM ($Y=1$). Notice that only the spin channels $(S,M)$ are relevant in the latter case. We have performed calculations at densities 0.16 and 0.08~fm$^{-3}$. The former value corresponds to the saturation density of SNM and the bulk density of finite nuclei. The latter one will give information about the surface of nuclei or the crust of neutron stars. All the displayed results have been calculated using the Skyrme interaction T44~\cite{les07}, as in previous works on SNM and PNM~\cite{dav09,pas12}. 

\subsection{Zero temperature}
\label{zeroT}
 
\begin{figure*}[!h]
\begin{center}
\includegraphics[width=0.43\textwidth,angle=-90]{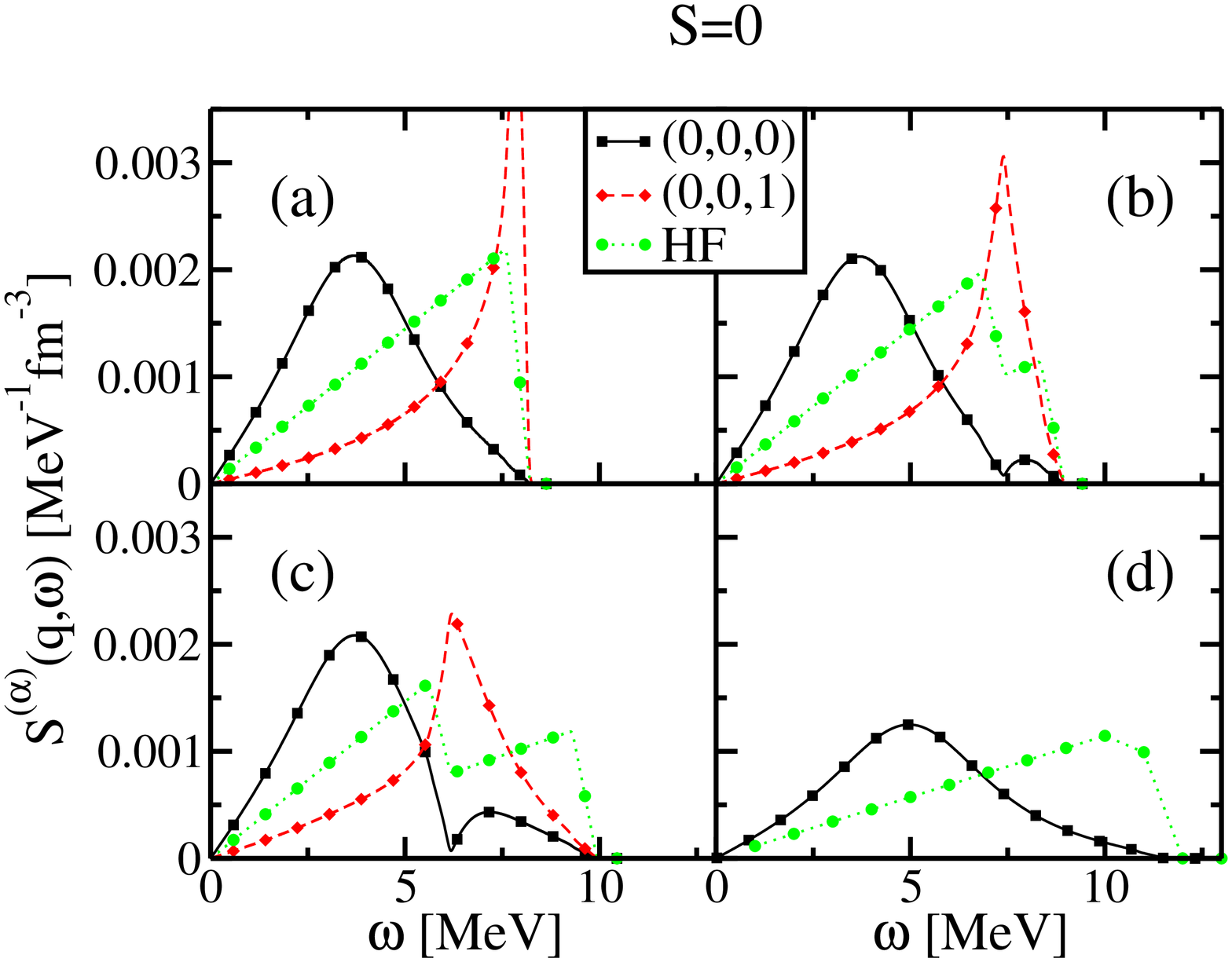}
\hspace{-2.9cm}
\includegraphics[width=0.43\textwidth,angle=-90]{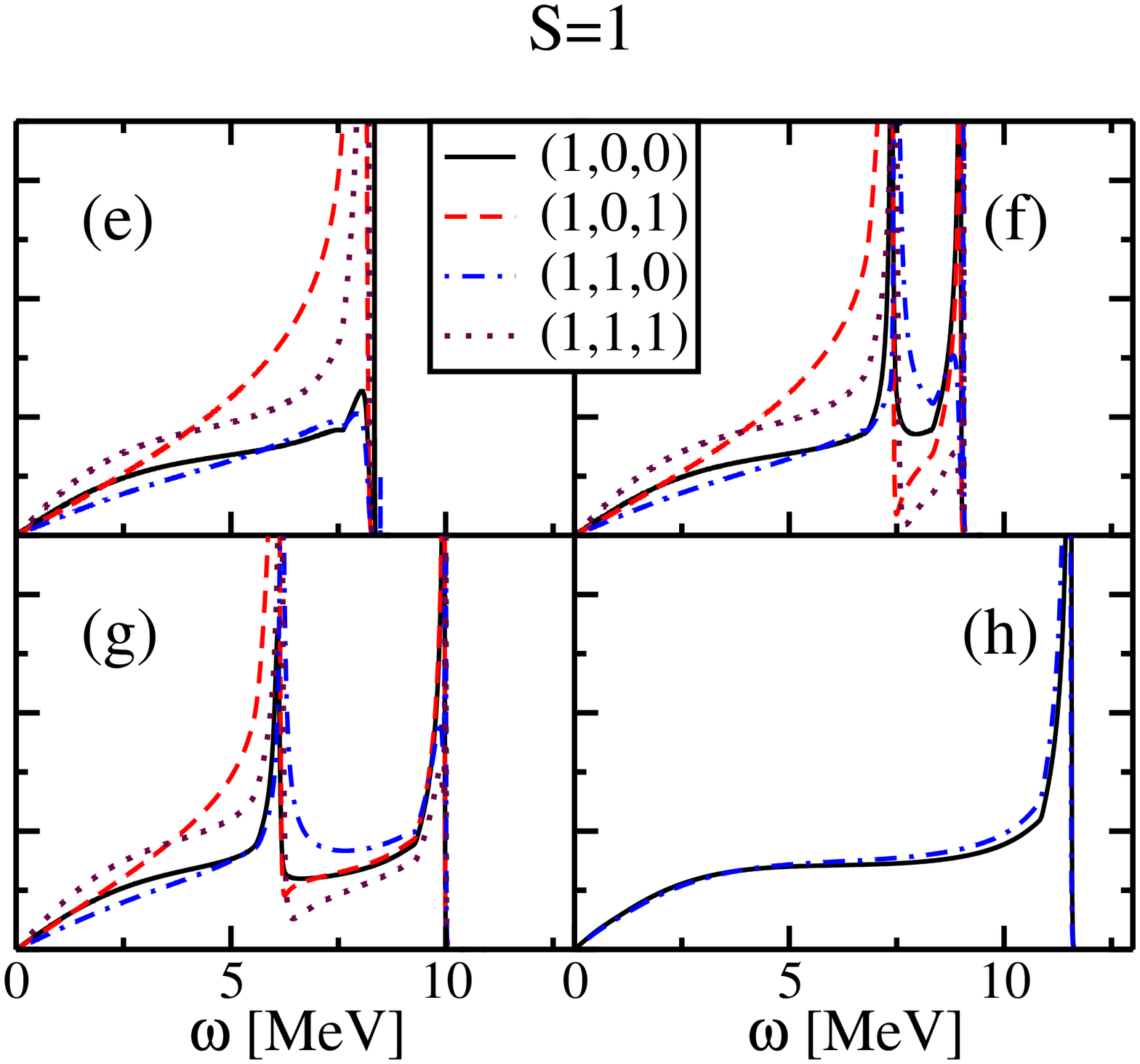}
\end{center}
\caption{(Color online) Strength functions $S^{(\alpha)}(q,\omega)$ for the spin-isospin channels $(\alpha)=(S,M,I)$ in asymmetric nuclear matter at density $\rho=0.16$ fm$^{-3}$ and momentum transfer $q=0.1$~fm$^{-1}$, calculated with Skyrme functional T44.  Panels (a) and (e) correspond to the asymmetry parameter $Y=1$ (SNM). Panels (b) and (f) to $Y=0.21$. Panels (c) and (g) to $Y=0.5$. Panels (d) and (h) to $Y=1$ (PNM). Only channels $(S,M)$ are relevant in PNM. The dotted lines in panels (a-d) are the $S_{HF}$ strengths.} 
\label{response:q01}
\end{figure*}

In Fig.~\ref{response:q01}, we show the strength functions $S^{(S,M,I)}(q,\omega)$ calculated at density $\rho= 0.16$~fm$^{-3}$ for a momentum transfer $q=0.1$~fm$^{-1}$. The spin channels $S=0, 1$ are displayed in panels (a-d) and (e-h), respectively. As a reference, the HF strength function --which is independent of the spin-isospin channel-- has also been displayed in panels (a-d). It can be seen that for the intermediate values of the asymmetry parameters ($Y=0.21$ and 0.50) the function $S_{HF}$ is in fact the superposition of two strengths, one for protons and the other for neutrons. Each one reflects the different Fermi momenta and effective masses of protons and neutrons. The residual ph interaction preserves this two-peaks structure in the RPA strength, except in channel $(S=0, I=0)$, which displays a single and broad resonance. The collective peak in the $(S=0, I=1)$ channel of SNM is shifted to the low-energy region as $Y$ increases, also reducing its height.  Regarding the $S=1$ channel, one can see that the collective states existing in both SNM and PNM is split in two well-separated peaks.  Their location is nearly the same for any value of $M$ and $I$. The tensor interaction manifests in particular as differences in the $M=0$ and 1 strength functions. These differences are more visible in the isospin $I=1$ channel and practically negligible in the $I=0$ one. 

\begin{figure*}[!h]
\begin{center}
\includegraphics[width=0.43\textwidth,angle=-90]{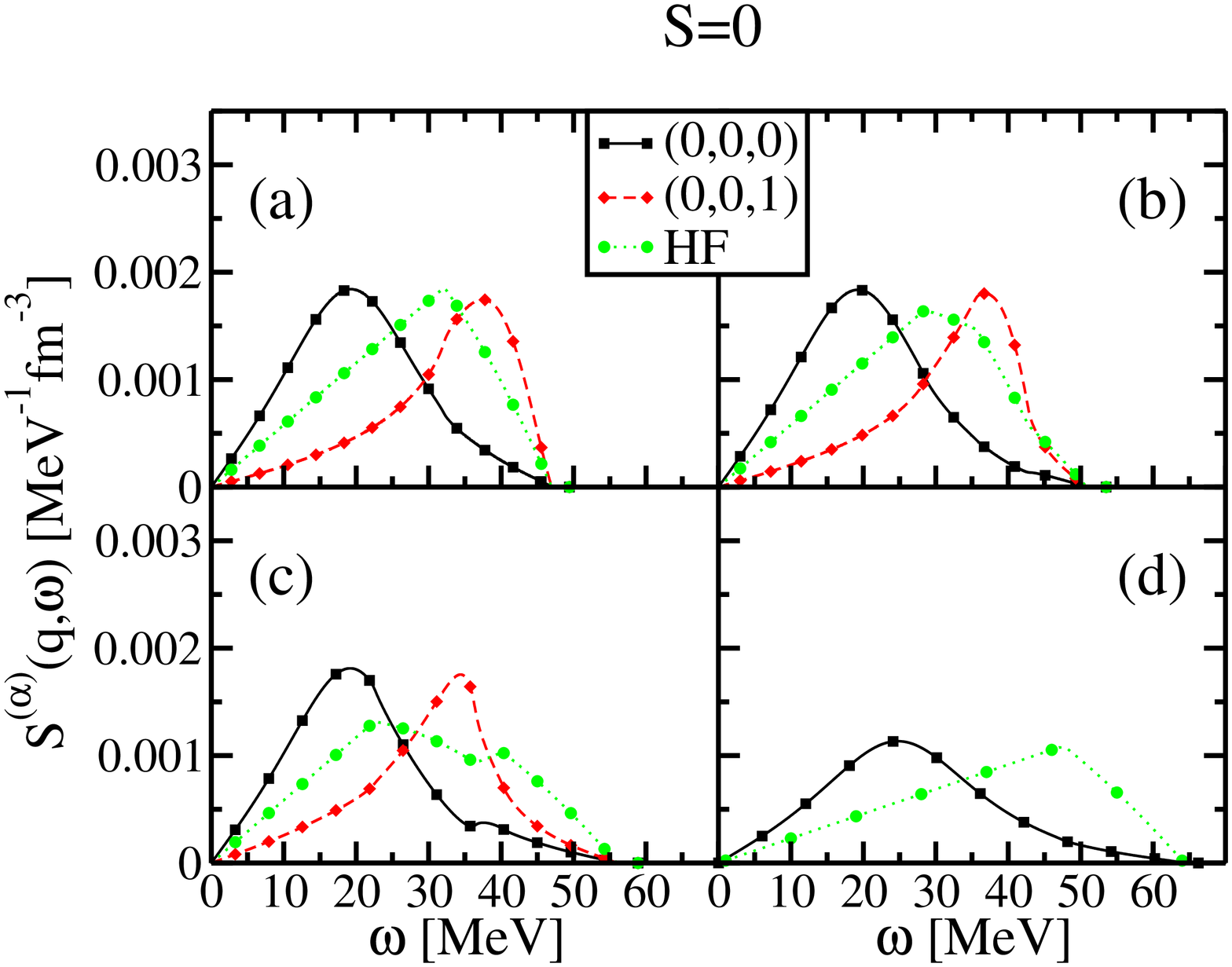}
\hspace{-2.9cm}
\includegraphics[width=0.43\textwidth,angle=-90]{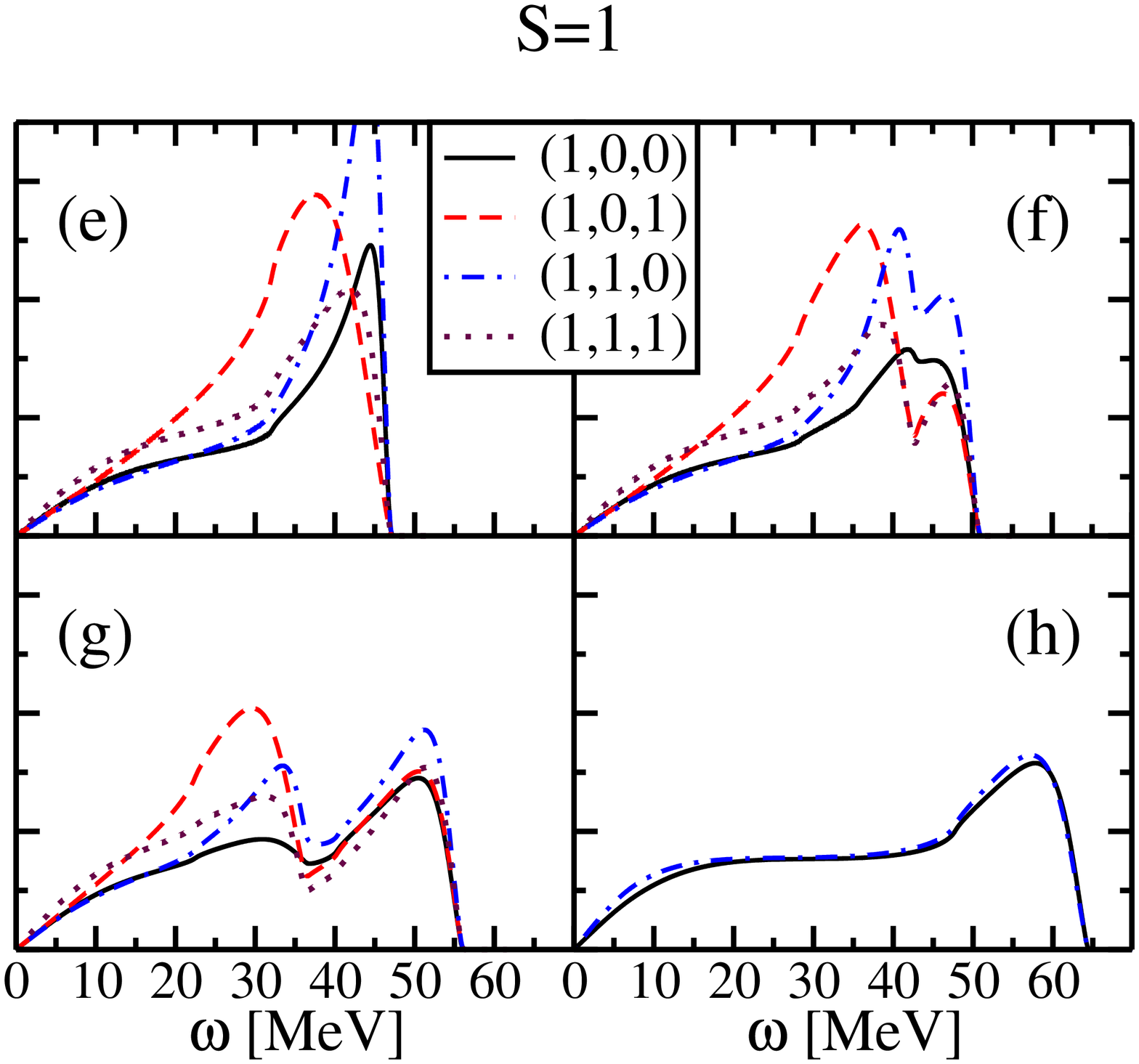}
\end{center}
\caption{(Color online)  Same as Fig.~\ref{response:q01}, for $q=0.5$~fm$^{-1}$.}
\label{response:q04}
\end{figure*}

\begin{figure*}[!h]
\begin{center}
\includegraphics[width=0.43\textwidth,angle=-90]{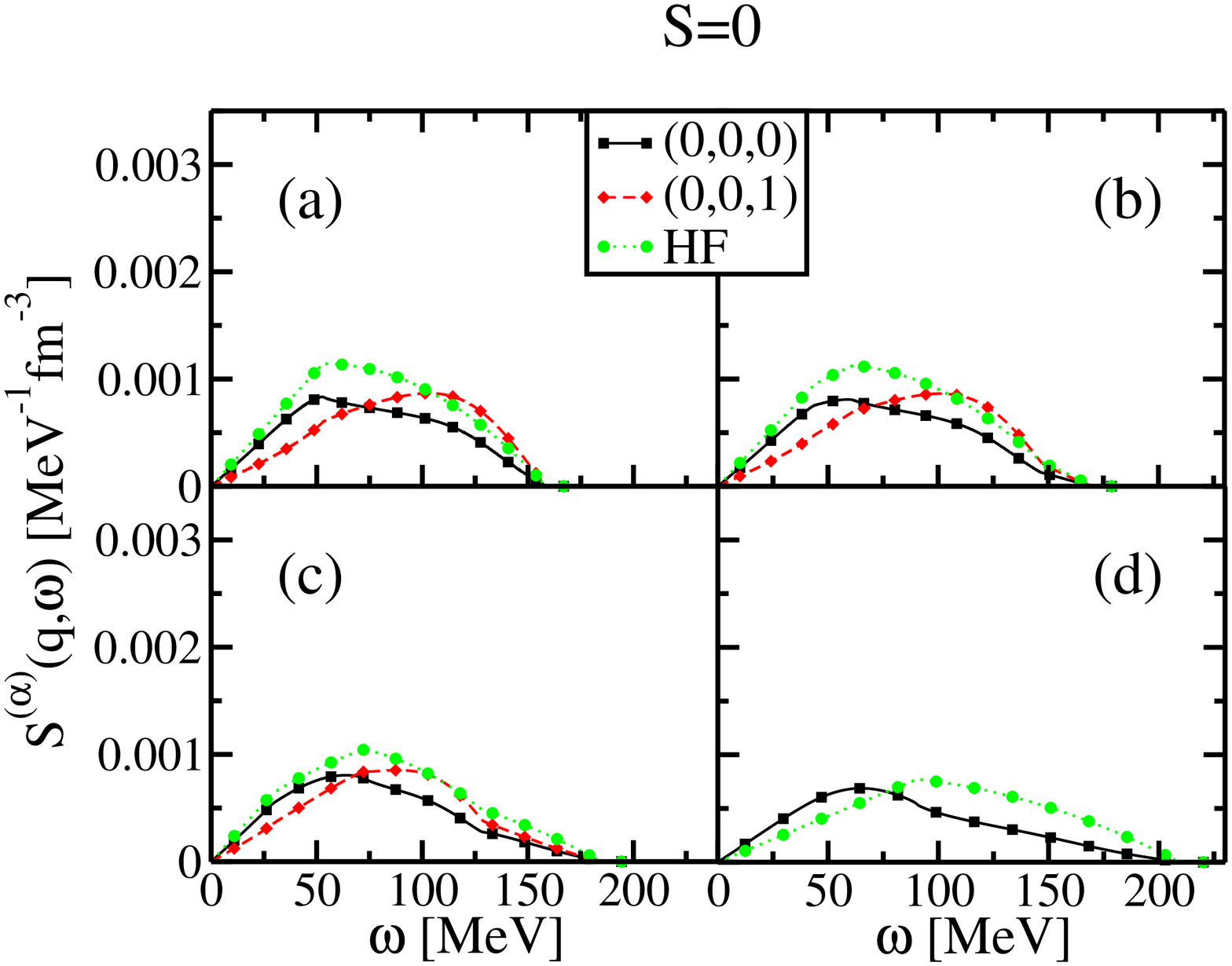}
\hspace{-2.9cm}
\includegraphics[width=0.43\textwidth,angle=-90]{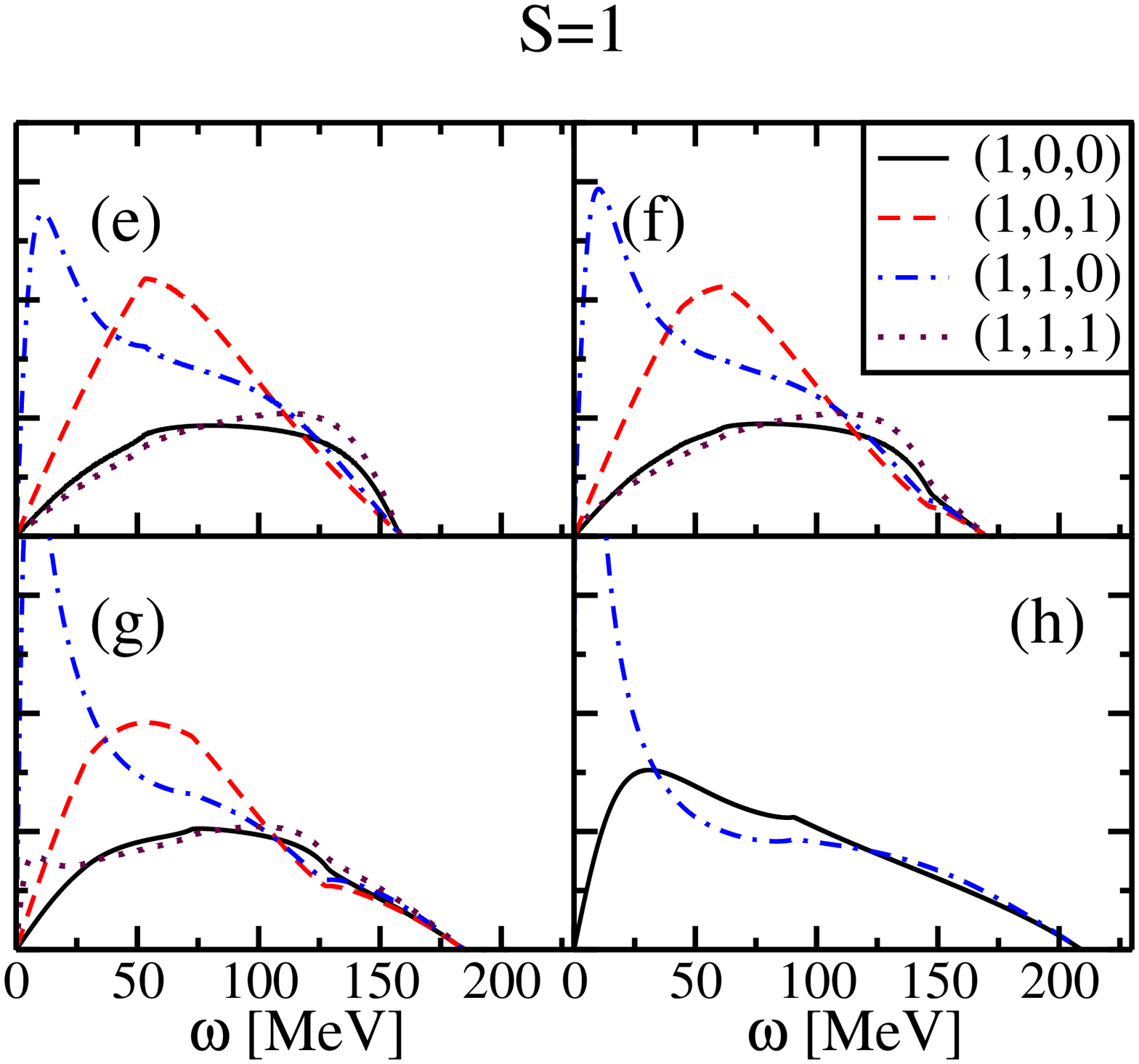}
\end{center}
\caption{(Color online)  Same as Fig.~\ref{response:q01}, for  $q=1.34$~fm$^{-1}$.}
\label{response:q1}
\end{figure*}

Strengths calculated at $q=0.5$ and 1.34~fm$^{-1}$ and the same density value are plotted in Figs.~\ref{response:q04} and~\ref{response:q1}, respectively. The two-peaks structure in the $S=1$ channel washes out as the momentum transfer is increased, becoming broader and less intense. Tensor effects are also magnified as $q$ increases, with more pronounced differences between the $M=0, 1$ strengths, irrespective of the isospin $I$. It is worth noticing the huge peak in channel $(1,0,0)$ at low values of $\omega$ for all values of $Y$. As will be discussed in Subsection \ref{instab}, this peak evolves to an instability at zero energy~\cite{pas12}.   

\begin{figure*}[!h]
\begin{center}
\includegraphics[width=0.43\textwidth,angle=-90]{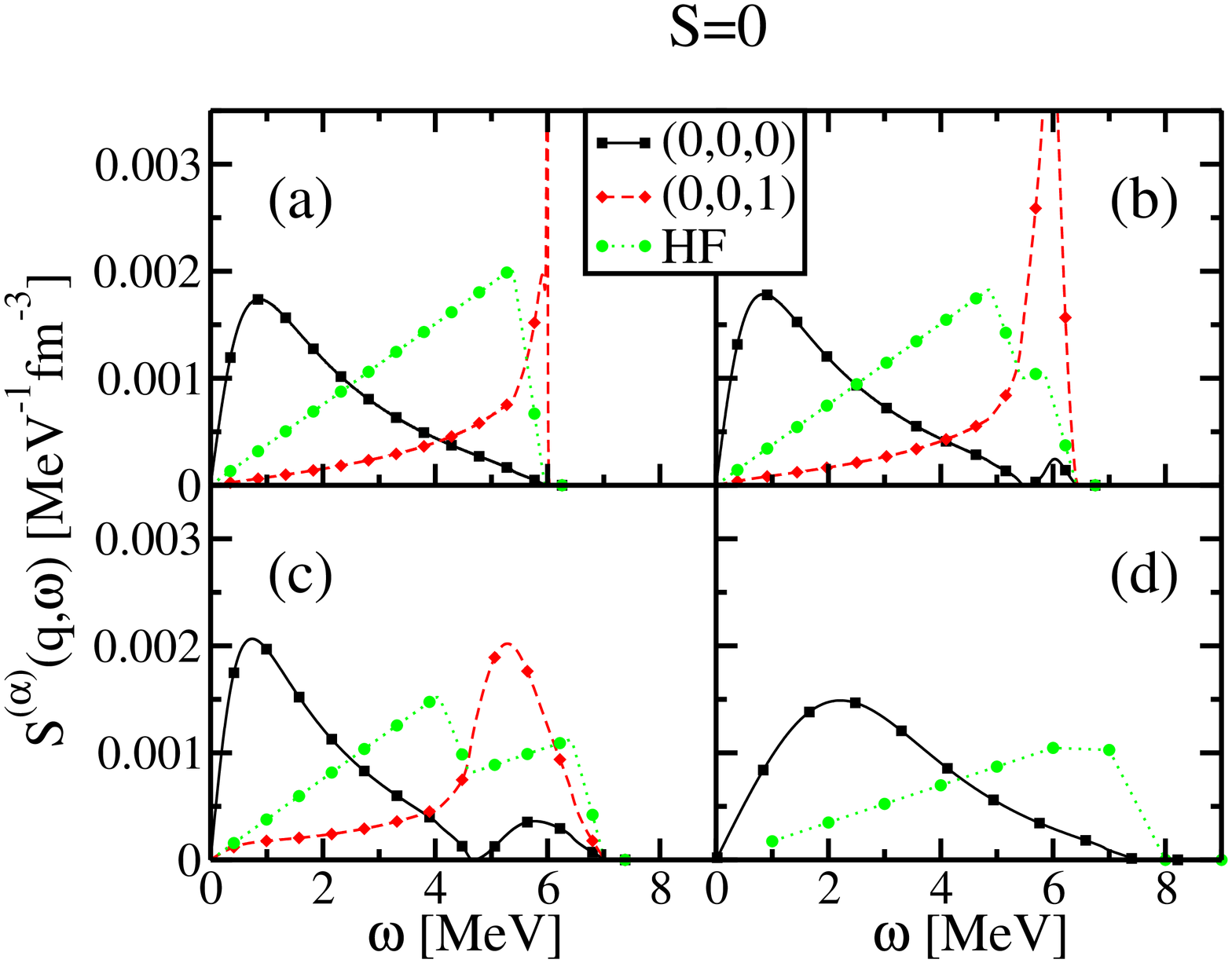}
\hspace{-2.9cm}
\includegraphics[width=0.43\textwidth,angle=-90]{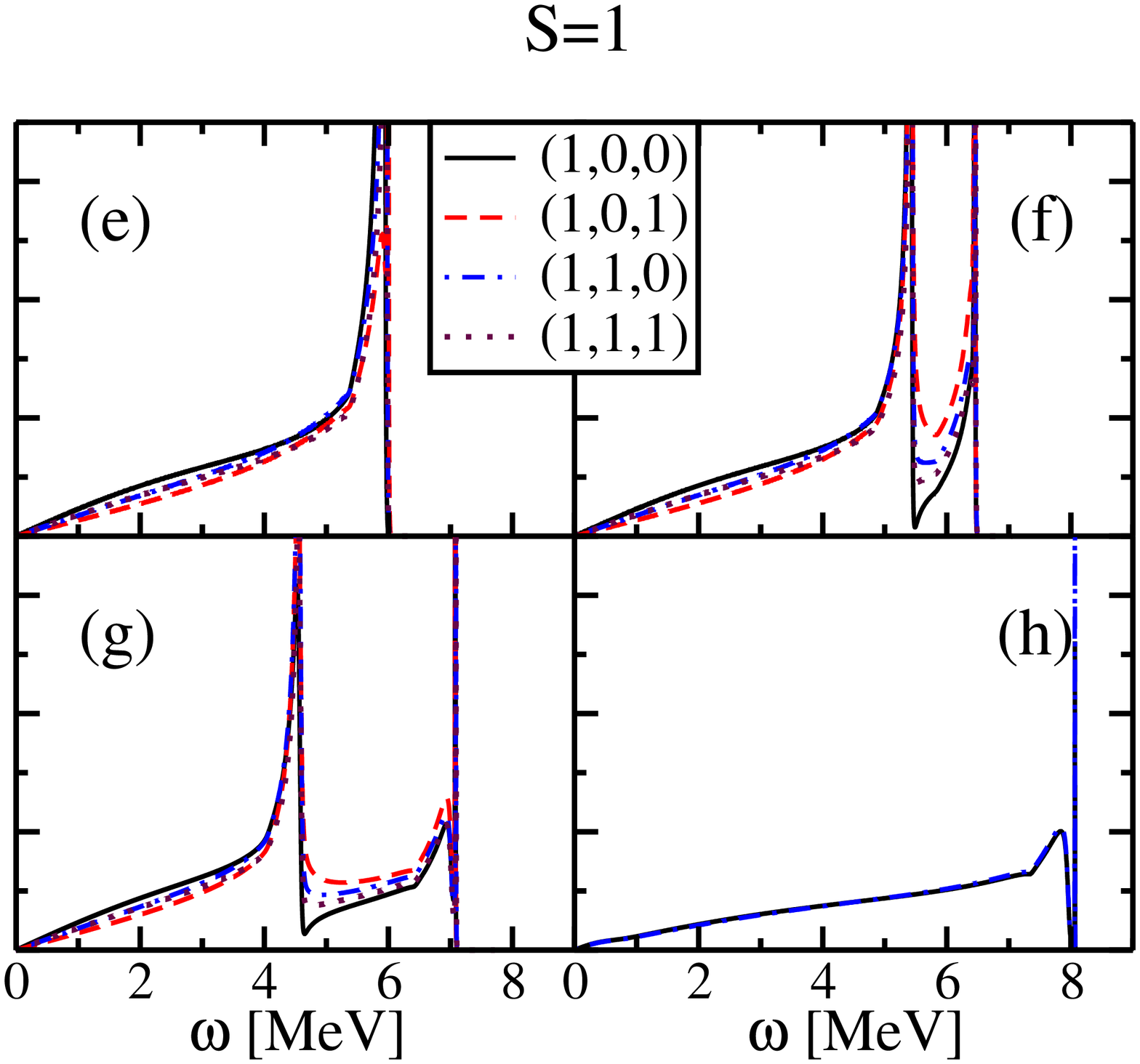}
\end{center}
\caption{(Color online) Same as Fig.~\ref{response:q01} for $\rho=0.08$ fm$^{-3}$.}
\label{response:q0108}
\end{figure*}

\begin{figure*}[!h]
\begin{center}
\includegraphics[width=0.43\textwidth,angle=-90]{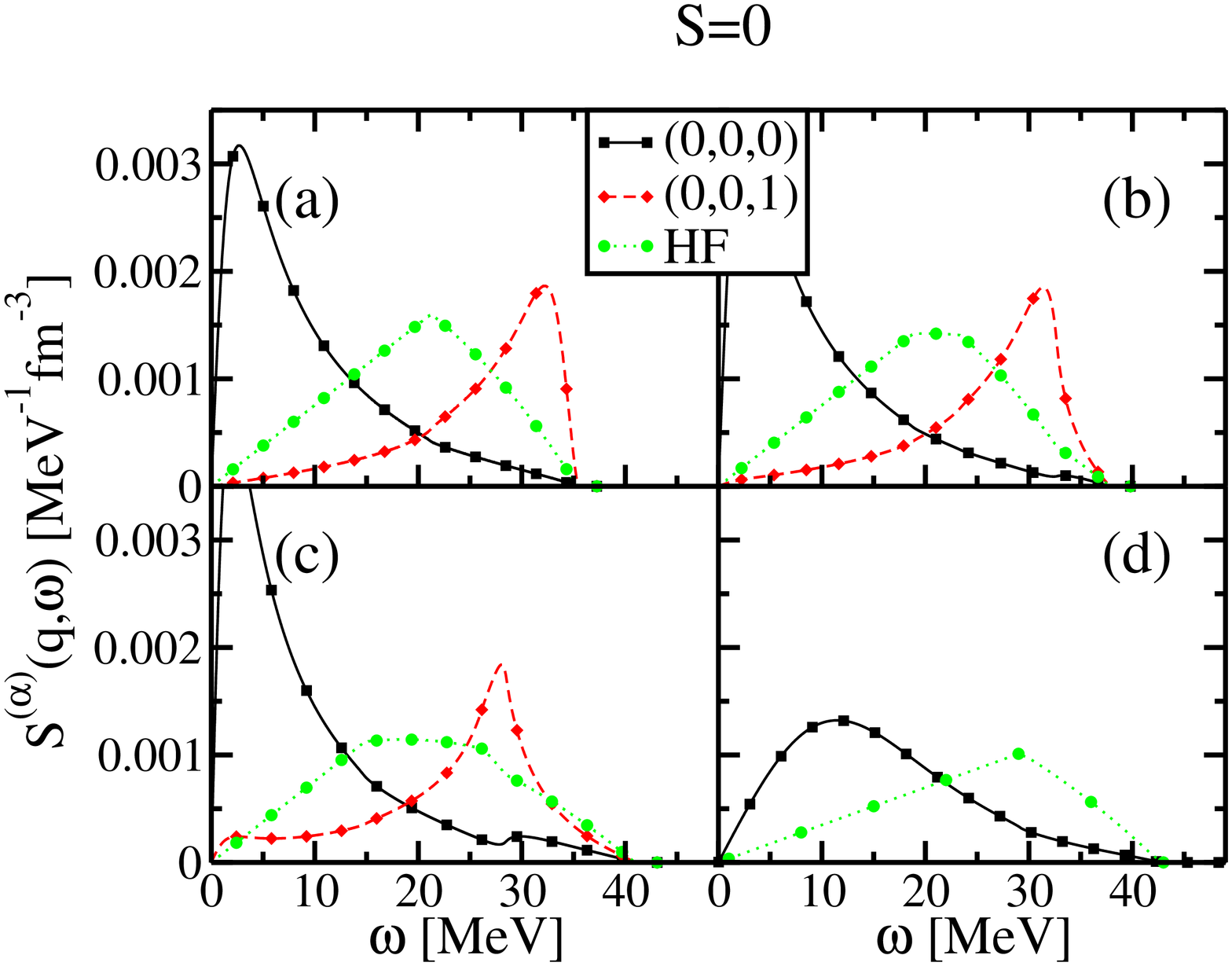}
\hspace{-2.9cm}
\includegraphics[width=0.43\textwidth,angle=-90]{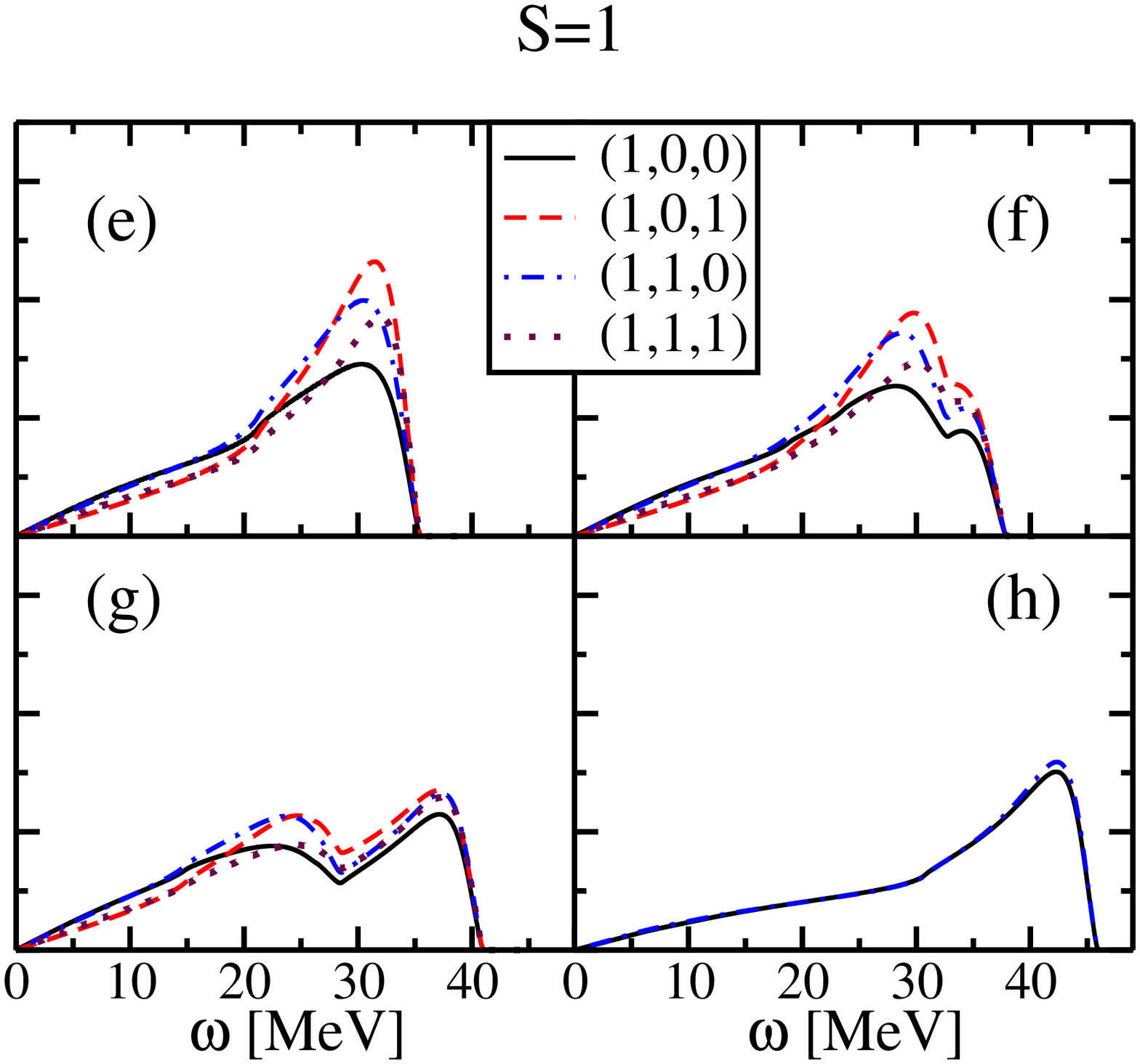}
\end{center}
\caption{(Color online)  Same as Fig.~\ref{response:q0108}, for  $q=0.5$~fm$^{-1}$. }
\label{response:q108}
\end{figure*}

Consider now the lower density $\rho=0.08$~fm$^{-3}$. Figs.~\ref{response:q0108} and \ref{response:q108} display the strengths calculated at the same transferred momenta than Figs.~\ref{response:q01} and \ref{response:q1}, respectively. A glance to the employed energy scales suffices to notice that the lowering of the density induces a global shift of the strength towards the low-energy region. For the transferred momentum $q=0.1$~fm$^{-1}$ the strengths in the different $S=1$ channels are nearly superimposed. For the asymmetries $Y=0.5$ and 1, a collective state is clearly visible in the channel $I=1$, well separated from the continuum edge. Analogously to Figs.~\ref{response:q01} and \ref{response:q1}, the strength function spreads as $q$ increases. However, an increase of the $(S=0, I=0)$ strength at low energies is noticeable. Actually,  as we shall discuss in Subsection \ref{instab}, this huge peak is related to the spinodal instability.

As already discussed in ref.~\cite{gar92,pas12} a very efficient tool to check the calculated linear responses is provided by their energy weighted sum rules. The sum rule or order $p$ is defined as 
\begin{eqnarray}\label{mmp:numeric}
M_p^{(S,M,I)}/A=-\frac{1}{\pi \rho}\int_0^{\infty} d\omega \, \omega^{p} \, \chi^{(S,M,I)}(q,\omega) \, ,
\end{eqnarray}
and can be numerically calculated from this expression. 
Alternatively, odd-order sum rules can also be obtained from appropriate expansions of response functions in power series of the transferred energy~\cite{boh79,lip89} as
\begin{eqnarray}\label{mmp:analytic}
\chi^{(S,M,I)}(\omega,q) \bigg|_{\omega \to \infty} &=& 2 \, \rho\sum_{p=0}^{\infty} \omega^{-(2p+2)}M^{(S,M,I)}_{2p+1}(q)/A \, , \\ 
\label{mm1:analytic}
\chi^{(S,M,I)}(\omega,q) \bigg|_{ \omega \to 0} &=& -2 \, \rho\sum_{p=0}^{\infty} \omega^{2p}M^{(S,M,I)}_{-(2p-1)}(q)/A \, .
\end{eqnarray}
The linear and cubic energy-weighted sum rules $M_1$ and $M_3$ can be easily obtained from the first two terms of (\ref{mmp:analytic}), while the inverse energy-weighted sum rule $M_{-1}$ is the first term of (\ref{mm1:analytic}). We have derived these analytic expressions using symbolic programs, but the expressions are too long to be presented here. For $M_{1}$, we have used the independent check of double-commutator technique~\cite{mey13} to properly verify our results. Our aim in comparing numerical and analytical sum rules is to have a good test about the reliability of our calculations. Besides, this comparison provides an independent way to localize collective states. 

\begin{figure}[!h]
\begin{center}
\includegraphics[width=0.5\textwidth,angle=-90]{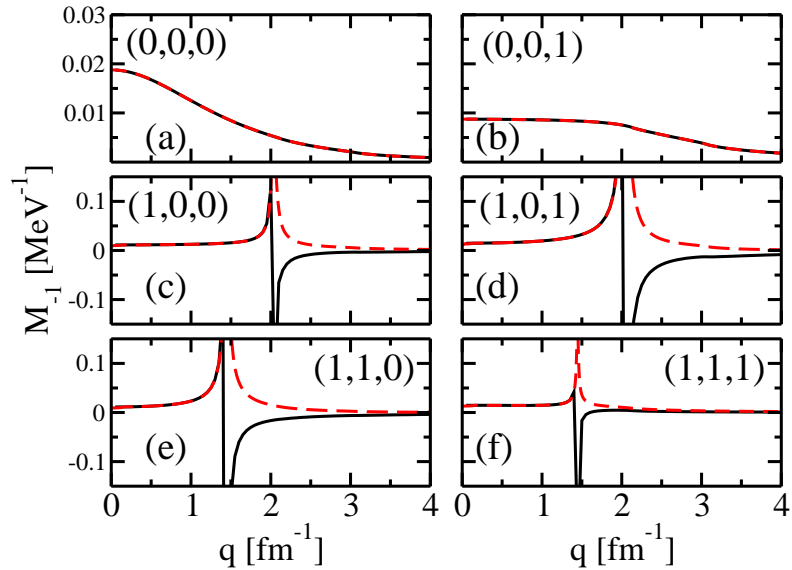}
\end{center}
\caption{(Color online) Inverse-energy weighted sum rule $M_{-1}$ for spin-isospin channels $(S,M,I)$ at asymmetry $Y=0.5$, as calculated analytically (solid line) and numerically (dashed line).}
\label{mm1:t44}
\end{figure}

The $M_{-1}$ sum rule is also interesting by itself because, apart from a trivial factor, it is the static susceptibility, or response function at zero energy. It is thus a very efficient tool to detect the instabilities related to a zero-energy mode, as already found in the case of SNM and PNM in~\cite{pas12,pas12b}.
Such a mode corresponds to a solution of $1/\chi^{S,M,I}(\omega=0,q)=0$, and has an infinite strength. Equivalently, these modes can be seen as divergencies in the sum rule $M_{-1}$. As an illustrative example, in Fig.~\ref{mm1:t44}, we present the results for the asymmetry parameter $Y=0.5$. Numerical (\ref{mmp:numeric}) and analytical (\ref{mm1:analytic}) sum rules $M_{-1}$ have been plotted as a function of $q$. For the spin $S=0$ channel the two curves stay on top of each other showing that the calculations are reliable. Both curves also coincide for $S=1$, except near some particular value of $q$, where a singularity appears. This kind of singularity will be discussed in Subsection \ref{instab}.

\subsection{Thermal effects}
\label{thermal}

We now consider the effect of temperature in the RPA response function. We refer the reader to Ref.~\cite{vau96} for a study about the inclusion of temperature in the many-body problem. Thermal effects in the RPA strength for SNM have been discussed in~\cite{bra95,her96} in the context of Skyrme interactions. In practice one has to modify the occupation probability of levels as 
\begin{equation*}
n_{\tau}(\mathbf{k})=\left\{ e^{(\varepsilon_{\tau}(\mathbf{k})-\mu_{\tau})/T}+1\right\}^{-1}
\end{equation*}
where $\mu_{\tau}$ is the chemical potential and $\varepsilon_{\tau}(\mathbf{k})$ is the HF single particle energy. 
The generalization to finite temperature of the Bethe-Salpeter equation (\ref{BSeq}) and the related algebraic systems of equations can be easily done. From a technical viewpoint one has simply to include it in the auxiliary functions $\beta^{\tau}_{i=0,8}(q,\omega,T)$ defined in Appendix~\ref{app:EQS}. 

At temperature $T=0$, the system is initially in the ground state, and the sole possible effect of the external 
probe is to excite the system, {\em i.e.} $\omega \ge 0$. 
However at non-null temperature the ground state of the system at equilibrium corresponds to a statistical 
mixture of states. Using the detailed-balance theorem, the strength function is properly defined as 
\begin{equation}
S^{(S,M,I)}(q,\omega,T) = -\frac{1}{\pi} \frac{{\rm Im} \chi^{(S,M,I)}(q,\omega,T)}{ 1-e^{\omega/T}}.
\end{equation}
In that case, it is possible to transfer energy from the system to the probe, so that negative values 
of $\omega$ are admissible.

\begin{figure}[!h]
\begin{center}
\includegraphics[width=0.38\textwidth,angle=-90]{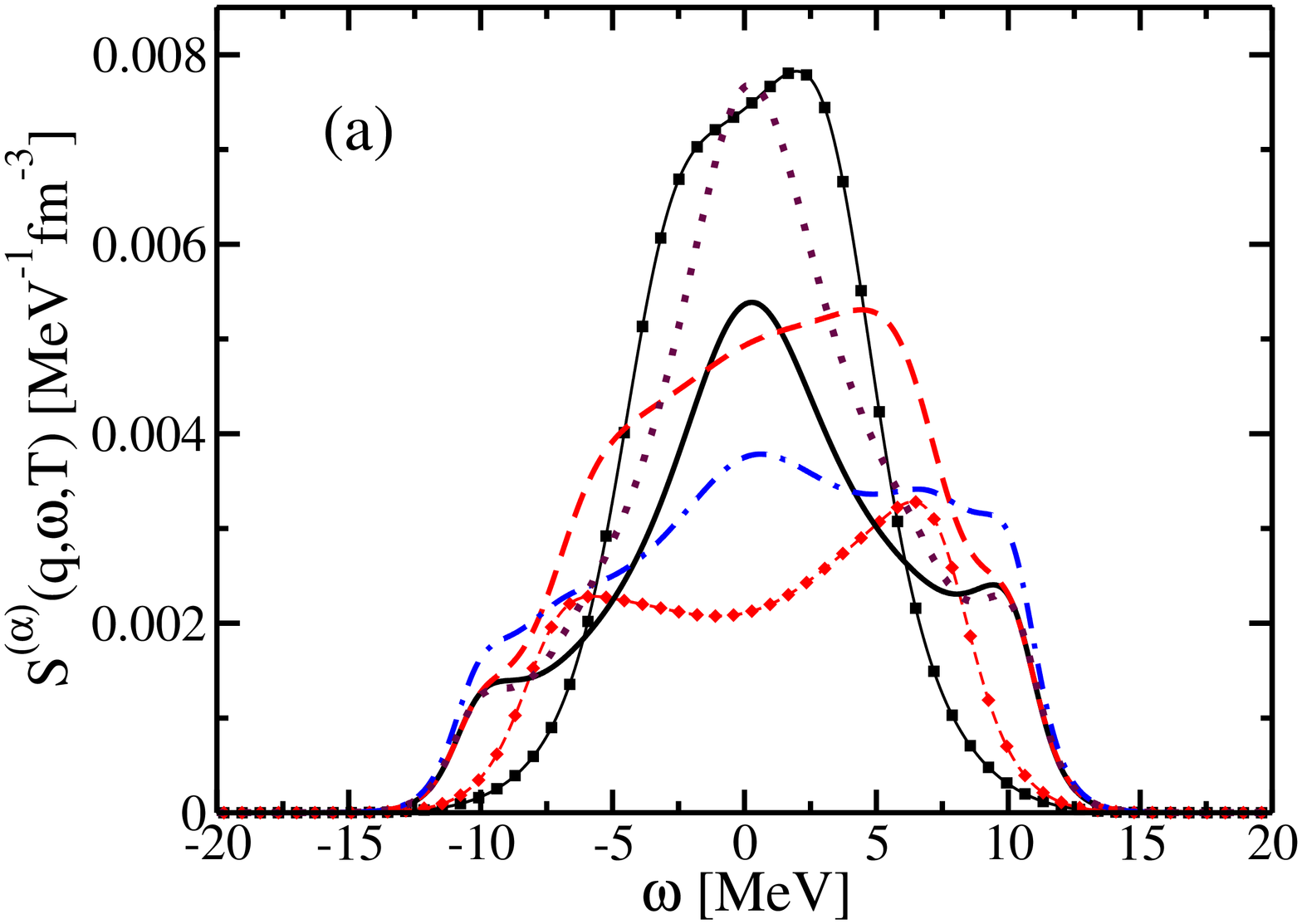}
\hspace{-2.76cm}
\includegraphics[width=0.38\textwidth,angle=-90]{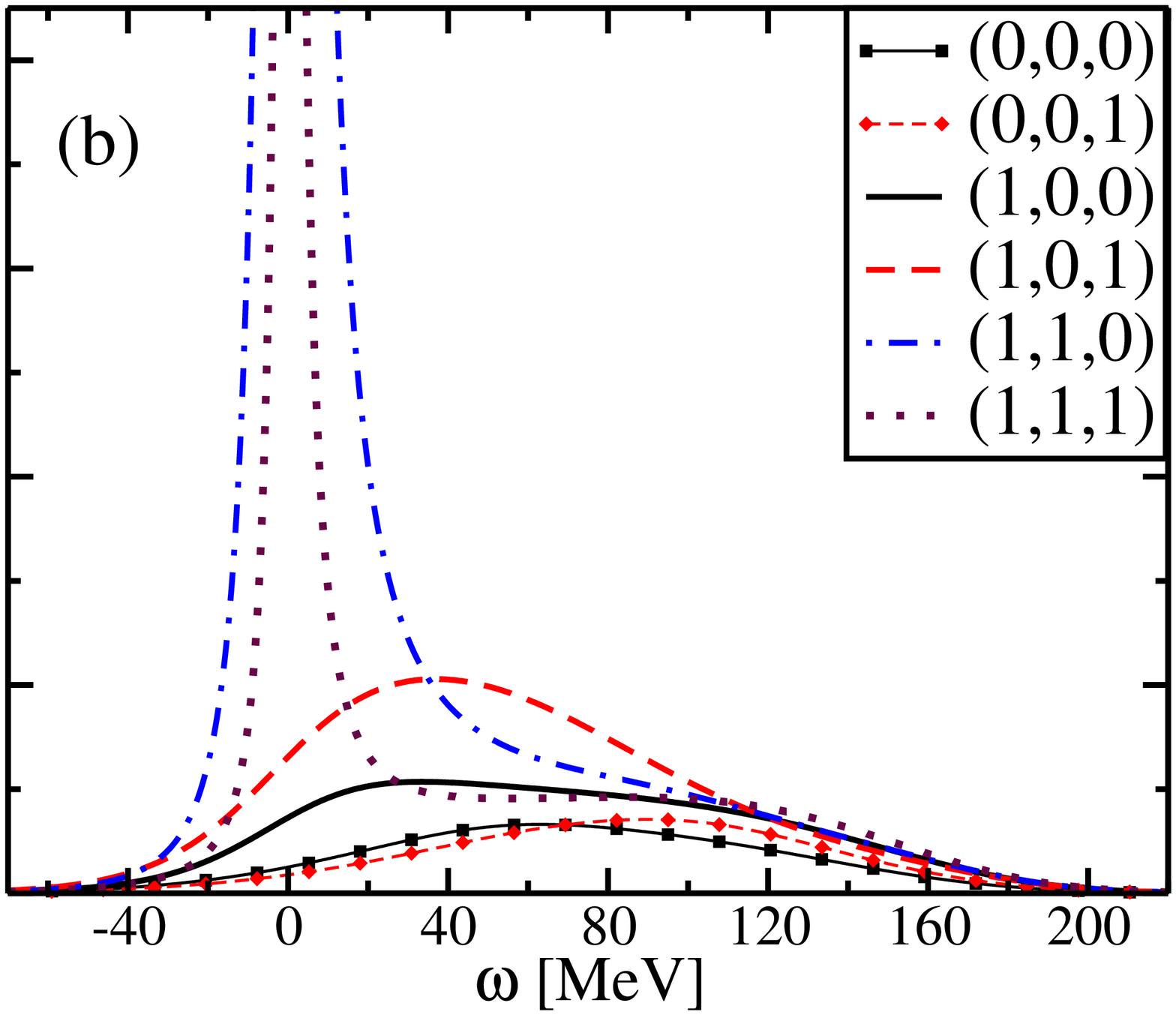}
\end{center}
\caption{(Color online) Strength function $S^{(\alpha)}(q,\omega,T)$ for spin-isospin channels $(\alpha)=(S,M,I)$ calculated at density $\rho=0.16$~fm$^{-3}$, temperature $T=16$~MeV and momentum transfer $q=0.1$~fm$^{-1}$ (left panel) and $q=1.34$~fm$^{-1}$ (right panel).}
\label{response:tt}
\end{figure}

Actually, thermal effects become relevant at values of temperatures larger than $\simeq 3$~MeV, and the limit of evaporation in finite nuclei appears at $T \approx 5$-10~MeV~\cite{bon84,bon85}. As a case of study we have performed calculations at $T=16$~MeV, a temperature which can be relevant for homogeneous systems of astrophysical interest. 
In Fig. \ref{response:tt} are displayed the strength functions calculated at density $\rho=0.16$~fm$^{-3}$ and asymmetry parameter $Y=0.5$, for two values of the transferred momentum, namely $q=0.1$~fm$^{-1}$ (left panel) and 1.34~fm$^{-1}$ (right panel). These responses should be compared to those in panels (c) and (g) of 
Figs.~\ref{response:q01} and \ref{response:q1}, respectively.  As a rule, temperature tends to wash out the structure of the response and spread its strength. An important part of the strength is shifted to the negative energy region, and the strength in some channels largely increases at zero energy. Notice also that each peak at positive energies has a corresponding ``image'' at negative energies, which corresponds to the deexcitation of the heated system. For this specific value of density, the strength in channels $S=1, M=1$ develops a huge zero-energy peak. Actually, it is the precursor of an unphysical instability, as we will immediately discuss.

\subsection{Instabilities}
\label{instab}

In the previous subsections we have encountered some cases were the strength function is hugely peaked at zero energy. This peak will become a divergence for specific values of density and momentum transfer. Some of these instabilities are unphysical, as they are simply reflecting drawbacks of the employed interaction~\cite{les06,hel13}. But at low values of density, the instabilities are related to the physical phenomenon of spinodal transition. 
As mentioned previously, the inverse-energy-weighted sum rule $M_{-1}$ is the tool of choice for the detection of poles of response functions at zero-transferred energy, and was employed in Refs~\cite{pas12,pas12b} to analyze instabilities in SNM and PNM.  

Let us consider first the instabilities appearing in the channel $(S,I)=(0,0)$ at low values of density. They are related to the thermodynamic spinodal transition of homogenous matter, where density fluctuations induce a decrease of the total free energy and are thus amplified until a separation in two distinct stable phases, liquid and gas, is reached. It has been shown~\cite{duc08} that the curvature of the free energy contains two terms: one proportional to $q^2$ coming from density gradient terms of the energy functional, and another proportional to $1/q^2$ related to the Coulomb interaction. For small values of $q$ the energy cost due to Coulomb interaction dominates and implies a sensitive reduction of the spinodal region for proton-rich systems. Note that in the case of a neutron star where a background of electrons is present, the net effect of the Coulomb interaction in the residual interaction is to reduce the region of the spinodal as well~\cite{duc08,bal09}. 

\begin{figure}[!h]
\begin{center}
\includegraphics[width=0.35\textwidth,angle=-90]{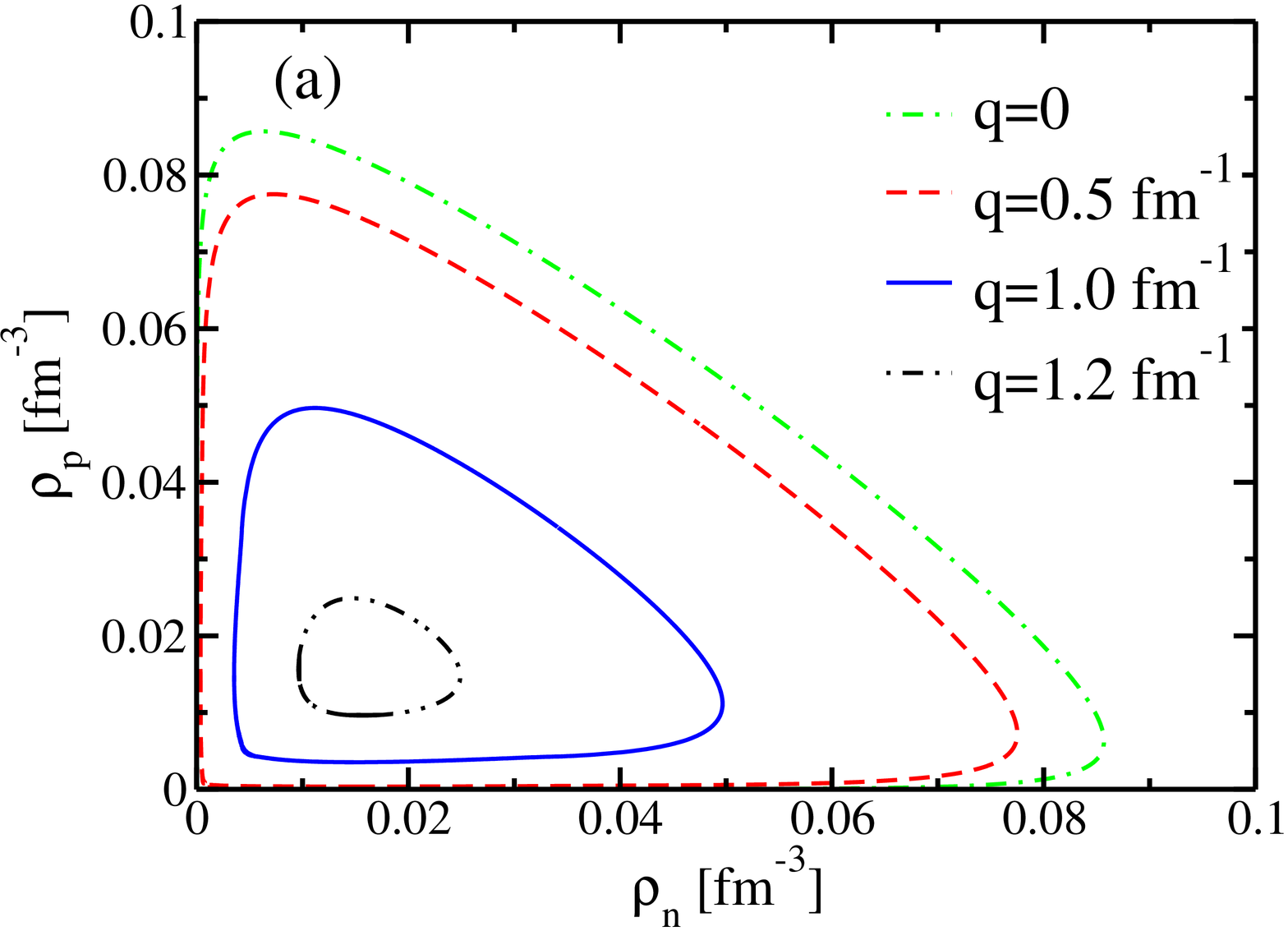}
\includegraphics[width=0.35\textwidth,angle=-90]{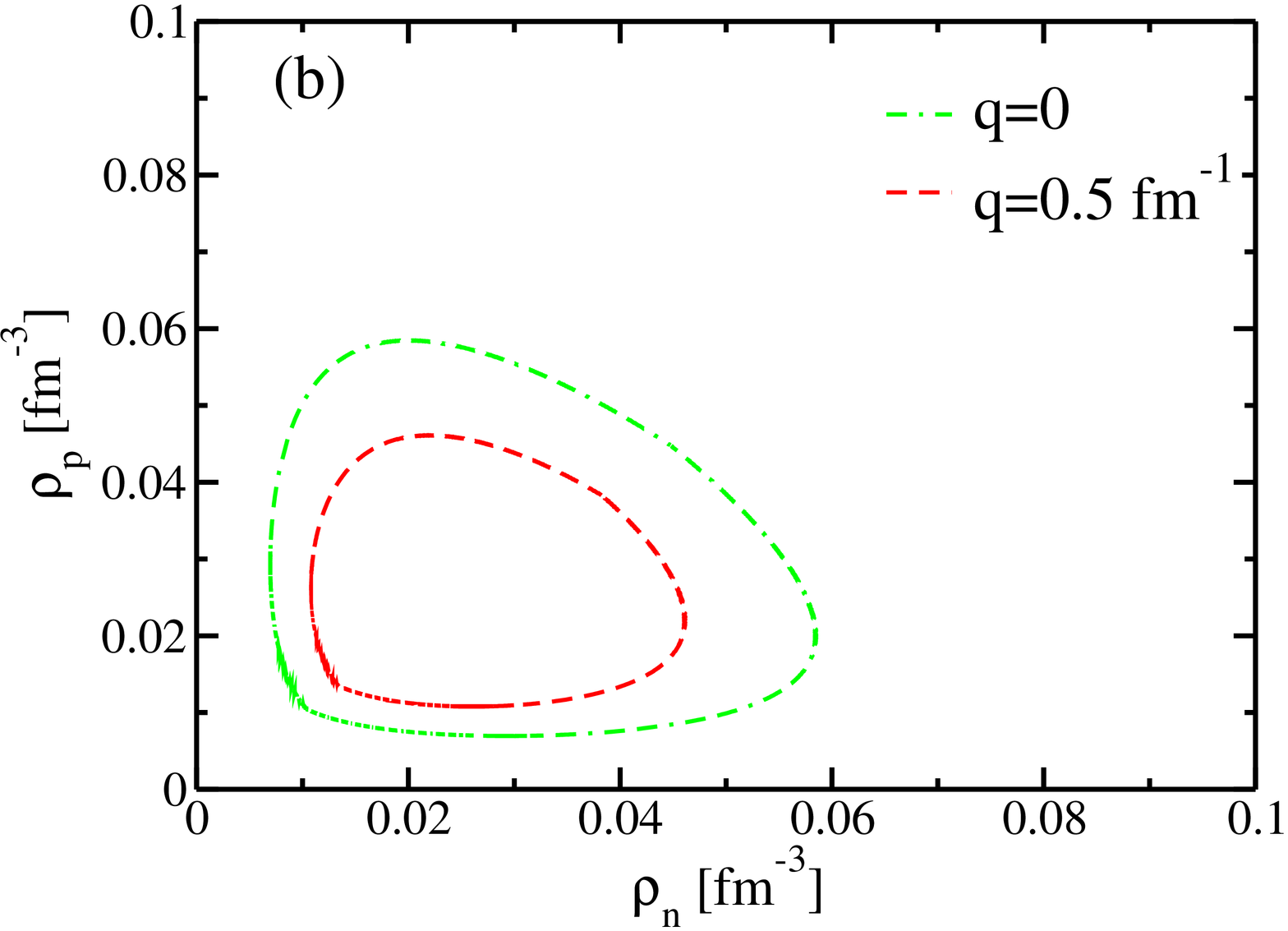}
\end{center}
\caption{(Color online) We show the spinodal instability (S=0,M=0,I=0) for the T44 Skyrme functional for different values of the neutron $\rho_{n}$ and proton $\rho_{p}$ density. On the left panel at T=0 and on the right panel for $T=10$~MeV.}
\label{spinodal:t44}
\end{figure}

In Fig.~\ref{spinodal:t44}, we have plotted the spinodal contours in the plane $(\rho_p,\rho_n)$ of proton/neutron densities for different values of the transferred momentum $q$ at zero temperature (left panel) and $T=10$~MeV (right panel). The homogeneous system is unstable inside these contours.  Since Coulomb interaction has been ignored in our description, the spinodal region is symmetric with respect to the line $\rho_p=\rho_n$, and is reduced as $q$ is increased, the largest possible region being that at $q=0$. In general, the effect of temperature is to suppress fluctuations. This can be seen on Fig.~\ref{spinodal:t44} by comparing the spinodal contours in both panels for a given value of $q$. Finally one may wonder about the influence of the tensor interaction on these instabilities: as we have mentioned in Sect.~\ref{sec:form}, although the tensor acts directly only in the $S=1$ channel, it can influence also the $S=0$ channel, due to the mixing between both spin channels induced by the spin-orbit term. This coupling has been absorbed into effective coefficients $\widetilde{W}_1^{(\tau \tau',S)}$ defined in Appendix~\ref{app-wp}). 
We have found that spin-orbit effects are very small, except at large values of the transferred momentum, as reflected in the $q^4$ power  entering explicitly in $\widetilde{W}_1^{(\tau \tau',S)}$.  As the spinodal instability concerns the $S=0$ channel and implies relatively small values of the transferred momentum, it is very marginally affected by the tensor interaction, which can be safely neglected to analyze the spinodal region.

\begin{figure*}[!h]
\begin{center}
\includegraphics[width=0.5\textwidth,angle=-90]{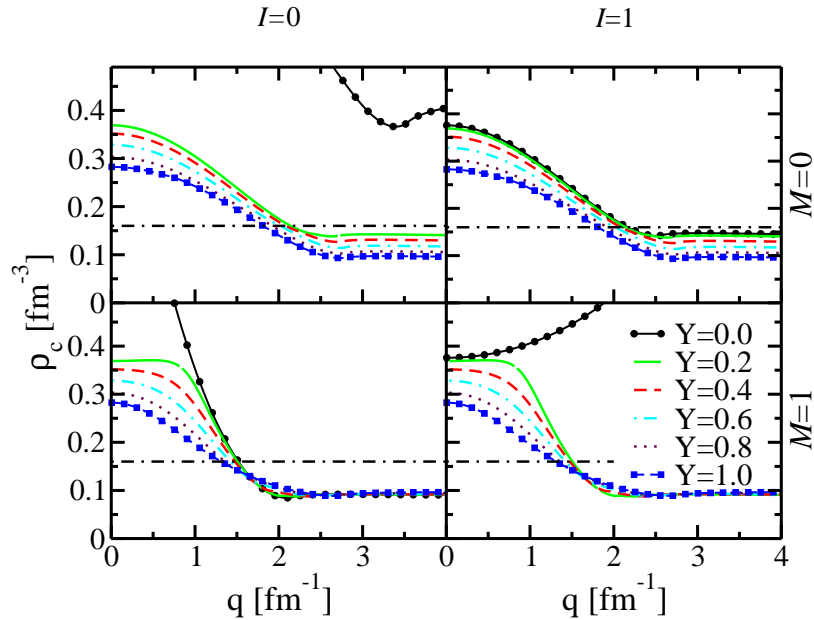}
\end{center}
\caption{(Color online)  In each panel we represent the critical density $\rho_{c}$ in the different channels $S=1,M=0,1,I=0,1$ for several values of the asymmetry parameter Y and at zero temperature. The horizontal dashed line corresponds to $\rho=0.16$~fm$^{-3}$.}
\label{s1:critical:t44}
\end{figure*}

We turn now to the unphysical instabilities, occurring in the $S=1$ channel. They have been analyzed in SNM and PNM in terms of Landau parameters associated to a specific Skyrme interaction~\cite{cao10,nav13,cha10,vid84,rio05,cha09}. As these instabilities are related to a specific interaction, fixing them will be of great help to establish bounds for the EDF coupling constants. For a given value of the transferred momentum $q$ we define a critical density $\rho_c$ as the value beyond which the homogeneous system becomes unstable. In Fig.~\ref{s1:critical:t44} are plotted the values of $\rho_c$ as a function of $q$ for $S=1$ channels and asymmetry parameters varying from $Y=0$ (SNM) to $Y=1$ (PNM) in steps of 0.2. Actually, for PNM there are only two channels, $(S=1,M)$, and the same curve is plotted in panels $I=0$ and $I=1$. One can see that for values of $q $ less that $\approx 1.2-1.4$~fm$^{-1}$, the critical density is higher than the saturation density of SNM (in the figure, the horizontal line corresponds to the value 0.16~fm$^{-3}$). The opposite happens for higher values of the transferred momentum. As a general trend, we can 
see that as a function of $Y$, the critical densities monotonically varies from a maximum value in SNM to a minimum value for PNM, with small variations around $q\approx 2$~fm$^{-1}$. However the decrease is not linear and might be actually very pronounced as soon as a small asymmetry is introduced. We should keep in mind that all the curves are actually obtained from a combination of response functions in the isospin channels $(nn)$, $(np)$, $(pn)$, and $(pp)$. Therefore, when a small asymmetry is introduced some of these response functions may suddenly acquire a non-zero value and influence a lot the whole combination. Although the calculations have been done with the single T44 interaction, one can extract a  pragmatic criterion: to use the bounds imposed by both SNM and PNM critical densities into a fitting procedure in order to get a stable interaction below such densities for all values of symmetries. We have also checked that the critical densities are reduced by only a few per cent at temperature $T=16$~MeV.

\section{Summary and conclusions}\label{sec:conclusion}

In this work we have generalized the formalism presented in Refs.~\cite{gar92,mar06,dav09} to calculate RPA response functions in asymmetric nuclear matter for a general Skyrme energy density functional including spin-orbit and tensor terms. The responses are obtained by solving closed algebraic systems of equations, which have been explicitly presented. Analytical expressions for the energy-weighted sum rules $M_{-1}$, $M_1$, and $M_3$ in terms of interaction parameters are easily derived from them by using an algebraic code. However, since the number of equations is doubled as compared to the SNM and PNM cases, the analytical expressions are rather cumbersome and it is preferable to compute them numerically.

We have investigated the response functions in the different spin-isospin channels for specific conditions. To illustrate the general trend concerning the effect of asymmetry we have chosen four values of the parameter $Y$. They correspond to the $^{208}$Pb nucleus, asymmetric nuclear matter in $\beta$-equilibrium, SNM and PNM. We have chosen two values of the density characteristic of homogeneous matter at saturation and the surface of atomic nuclei. 
All calculations have been done using the T44 functional as a typical one, and so our conclusions are not completely general. However, it seems that the behavior of ANM is not just a simple interpolation between SNM and PNM even if one has clearly identified how to go from one limiting case to the other in an analytical way. 

We have found that the tensor plays an important role in the response, leading to the presence of two well-defined collective states. In particular, as compared to the HF response, the tensor significantly amplify the separation between the two Fermi surfaces. Even small values of asymmetry have a sensible effect on response functions. This is at variance with the results of Ref.~\cite{her97a} where it was concluded that varying the asymmetry parameter produces no spectacular variations in the response. However, these results were obtained using purely central forces, while important tensor effects have already been observed in our previous studies concerning SNM and PNM.
Concerning thermal effects, we have shown that up to temperatures of 16~MeV, the response function is not sizably modified. 

The static susceptibility is a key quantity to get some information about instabilities, and we have computed it through the related inverse energy-weighted sum rule. Our results concerning the spinodal instability compare favorably with those of~\cite{duc08}. The general behavior obtained here for the spinodal region with respect to asymmetry and temperature reproduce nicely the physical features demonstrated in a very different context, as is the study of free energy curvature in fluctuation space~\cite{duc08b}. Apart from this physical instability, we have also put into evidence the unphysical instabilities, thus generalizing to ANM the findings of~\cite{pas12,pas12b}. However it is not so simple to 
define finite-size instabilities and critical densities. Depending on the channel, the asymmetry induces an important modification of the location of the poles or a minor one. Moreover, the lowest pole is not obtained for the same asymmetry, even if PNM was generally favored. As a practical rule, the bounds imposed by PNM critical densities are sufficient as a criterion to get stable interactions within a standard fitting procedure.

\section*{Acknowledgements}
We would like to thank C. Ducoin and J. Meyer for interesting discussions and remarks about this paper. 
This work was partly supported by grant FIS2011-28617-C02-02, Mineco (Spain). 


\begin{appendix}
\section{Coefficients $W_a^{(\tau,\tau',S)}$ and EDF parameters}
\label{app:w}

We give here the complete expressions of the coefficients $W_a^{(\tau,\tau',S)}$ in terms of the coupling constant of the functional \cite{per04}. We distinguish the cases $\tau=\tau'$ (i.e. both indices equal to $p$ or $n$) and $\tau=-\tau'$ (i.e. indices equal to $p,n$ or $n,p$).
\begin{eqnarray}
W_1^{(\tau,\tau,S)} &=& W_1^{(S,0)} + W_1^{(S,1)} + b(\tau) 8 C_1^{\rho,\gamma} \rho_1 \gamma \rho^{\gamma-1}
+ W_{T2}^{(0)} + W_{T2}^{(1)} \,,\\
W_1^{(\tau,-\tau,S)} &=& W_1^{(S,0)} - W_1^{(S,1)} + W_{T2}^{(0)} - W_{T2}^{(1)}\,,
\end{eqnarray}
with $b(n)=1$ and $b(p)=-1$. For the remaining indices ($a= 2, T1, SO$) we have
\begin{eqnarray}
W_a^{(\tau,\tau,S)} &=& W_a^{(S,0)} + W_a^{(S,1)} \,, \\
W_a^{(\tau,-\tau,S)} &=& W_a^{(S,0)} - W_a^{(S,1)}\,.
\end{eqnarray}
The coefficients $W_a^{(S,T)}$ are
\begin{eqnarray}\label{Wfunction}
\frac{1}{4}W^{(0,0)}_1&=&2C_0^{\rho 0}+(2+\gamma)(1+\gamma)C_0^{\rho \gamma}\rho_0^{\gamma}+\gamma(\gamma-1) C^{\rho,\gamma_{}}_1\rho_0^{\gamma_{}-2}\rho_1^{2} -\left[ 2C_0^{\Delta \rho}+\frac{1}{2}C_0^{\tau}\right]q^{2}\\
\frac{1}{4}W^{(0,1)}_1&=&2C^{\rho 0}_1+2C^{\rho,\gamma_{}}_1\rho_0^{\gamma_{}} -\left[2 C_1^{\Delta \rho}+\frac{1}{2}C_1^{\tau}\right]q^{2}\,,\\
\frac{1}{4}W^{(1,0)}_1&=& 2C_0^{s,0}+2C_0^{s\gamma_{}}\rho_0^{\gamma_{}} -\left[2C_0^{\Delta s}+\frac{1}{2}C_0^{T} \right]q^{2}\,,\\
\frac{1}{4}W^{(1,1)}_1&=& 2C_1^{s,0}+2C_1^{s\gamma_{}}\rho_0^{\gamma_{}}- \left[2C_1^{\Delta s}+\frac{1}{2}C_1^{T} \right] q^{2}\,,\\
\frac{1}{4}W^{(0,0)}_2&=&C_0^{\tau}\,, \\
\frac{1}{4}W^{(0,1)}_2&=&C_1^{\tau}\,,\\
\frac{1}{4}W^{(1,0)}_2&=&C_0^{T}\,,\\
\frac{1}{4}W^{(1,1)}_2&=&C_1^{T}\,.
\end{eqnarray}
For vanishing isovector density, they coincide with those previously given for symmetric nuclear matter \cite{gar92,dav09}. The remaining coefficients are related to the tensor and spin-orbit components of the interaction
\begin{eqnarray}
W_{T1}^{(T)} &=& 4 C_T^F \,, \\
W_{T2}^{(T)} &=& 8 C_T^{\nabla s} - 2 C_T^F\,, \\
W_{SO}^{(T)} &=& 4 C_T^{\nabla J} \,.
\end{eqnarray}

We also define for convenience : $C_\pm^F \equiv C_0^F \pm C_1^F$ and $C_\pm^{\nabla J} \equiv C_0^{\nabla J} \pm C_1^{\nabla J}$.

\section{Algebraic system of equations}
\label{app:EQS}

Before giving the explicit expressions we have to define some averages of the HF ph propagator
\begin{equation}
\label{app:beta:betafunct}
\beta_{i}^{(\tau)}(q,\omega,T) =\int \frac{d^{3}k}{(2\pi)^{3}}G_{HF}^{(\tau)}(\mathbf{k},q,\omega)F_{i}(\mathbf{k},q)\,,
\end{equation}
where the functions $F_{i}(\mathbf{k},q)$ are defined as
\begin{equation}
F_{0,..,8}=1,\frac{\mathbf{k}\cdot q}{q^{2}},\frac{k^{2}}{q^{2}},\left[\frac{\mathbf{k}\cdot q}{q^{2}}\right]^{2},\frac{(\mathbf{k}\cdot q)k^{2}}{q^{4}},\frac{k^{4}}{q^{4}},\left[\frac{\mathbf{k}\cdot q}{q^{}}\right]^{3},\left[\frac{\mathbf{k}\cdot q}{q}\right]^{4},\frac{(\mathbf{k}\cdot q)^{2}k^{2}}{q^{4}}\,,\nonumber
\end{equation}
for $i=0$ to 8, respectively. The explicit expressions of these functions can be found in ref.\cite{her97a} up to $i=5$. The three remaining functions (for $i=6, 7, 8$) are required by the inclusion of tensor terms. Its generalization is straightforward and will not be given here. 
Obviously, for the extreme values $Y=0$ and $Y=1$ one recovers the expressions of the $\beta$ functions given in~\cite{dav09}.
For finite temperature, only the imaginary parts are determined explicitly. The real parts are obtained as usual through a dispersion relation.

We give now the explicit form of the column vectors $X_{nn}$, $X_{np}$, $B_n$ and the matrices $A_{nn}$, $A_{np}$. 
The blocks $A_{pp}$ and $A_{pn}$ are obtained from $A_{nn}$ and $A_{np}$ by simply exchanging $n \leftrightarrow p$. These quantities define completely the system (\ref{pepet}) and obtain $\langle G^{(nn)}_{RPA}\rangle$ and $\langle G^{(pn)}_{RPA}\rangle$. To determine $\langle G^{(pp)}_{RPA}\rangle$ and $\langle G^{(np)}_{RPA}\rangle$ one must replace  $n\leftrightarrow p$ in the previous expressions.
The coefficients $\widetilde{W}_1^{(\tau \tau',SM)}$ and $\alpha_i^{j \tau}$ entering the following expressions are defined in Appendix~\ref{app-wp}.

\subsection{Channel S=0}

\begin{eqnarray}\label{XB:S0}
X_{nn} =
\left( \begin{matrix}
\langle G^{(nn)}_{RPA}\rangle\\
\langle k^{2} G^{(nn)}_{RPA}\rangle\\
\sqrt{\frac{4\pi}{3}}\langle kY_{10}G^{(nn)}_{RPA}\rangle\\
\end{matrix}\right)  , \,
X_{pn} =
\left( \begin{matrix}
\langle G^{(pn)}_{RPA}\rangle\\
\langle k^{2}G^{(pn)}_{RPA}\rangle\\
\sqrt{\frac{4\pi}{3}}\langle kY_{10}G^{(pn)}_{RPA}\rangle\\
\end{matrix}\right)  , \,
B_n= \left( \begin{matrix}\beta_0^{(n)}\\
q^{2}\beta_2^{(n)}\\
q\beta_1^{(n)}
\end{matrix}\right)
\end{eqnarray}

\begin{eqnarray}
A_{nn} = \left( \begin{matrix}
1 - \beta_0^{(n)} \widetilde{W}_1^{(nn,0)} - q^2 \beta_2^{(n)} W_2^{(nn,0)}
&-\beta_0^{(n)} W_2^{(nn,0)}
&  2q \beta_1^{(n)} W_2^{(nn,0)} \\
-q^2 \beta_2^{(n)} \widetilde{W}_1^{(nn,0)} - q^4 \beta_5^{(n)} W_2^{(nn,0)}
&  1-q^2 \beta_2^{(n)} W_2^{(nn,0)}
&2q^3 \beta_4^{(n)} W_2^{(nn,0)} \\
-q \beta_1^{(n)} \widetilde{W}_1^{(nn,0)} - q^3 \beta_4^{(n)} W_2^{(nn,0)}
& -q \beta_1^{(n)} W_2^{(nn,0)}
&1 +2 q^2 \beta_3^{(n)} W_2^{(nn,0)} \\
\end{matrix} \right)
\end{eqnarray}
\begin{eqnarray}
A_{np} = \left( \begin{matrix}
-\beta_0^{(n)}\widetilde{W}_1^{(np,0)}-q^{2}\beta_2^{(n)}W_2^{(np,0)}
&  -\beta_0^{(n)}W_2^{(np,0)}
& 2q \beta_1^{(n)}W_2^{(np,0)}\\
-q^{2}\beta_2^{(n)}\widetilde{W}_1^{(np,0)}-q^{4}\beta_5^{(n)}W_2^{(np,0)}
& -q^{2}\beta_2^{(n)}W_2^{(np,0)}
&2q^{3}\beta_4^{(n)}W_2^{(np,0)}  \\
-q\beta_1^{(n)}\widetilde{W}_1^{(np,0)}-q^{3}\beta_4^{(n)}W_2^{(np,0)}
&-q\beta_1^{(n)} W_2^{(np,0)}
& 2q^{2}\beta_3^{(n)} W_2^{(np,0)}\\
\end{matrix}
\right)
\end{eqnarray}

\subsection{Channel S=1 M=$\pm$1}

\begin{eqnarray}
X_{nn}= \left( \begin{matrix}
\langle G^{nn}_{RPA}\rangle\\
\langle k^{2} G^{nn}_{RPA}\rangle\\
\sqrt{\frac{4\pi}{3}}\langle kY_{10}G^{nn}_{RPA}\rangle\\
\frac{8\pi}{3}\langle k^2|Y_{11}|^2G^{nn}_{RPA}\rangle\\
\end{matrix}\right) \, , \,
X_{pn}= \left( \begin{matrix}
\langle G^{pn}_{RPA}\rangle\\
\langle k^{2}G^{pn}_{RPA}\rangle\\
\sqrt{\frac{4\pi}{3}}\langle kY_{10}G^{pn}_{RPA}\rangle\\
\frac{8\pi}{3}\langle k^2|Y_{11}|^2G^{pn}_{RPA}\rangle\\
\end{matrix}\right) \, , \,
B_n =  \left(
\begin{matrix}
\beta_0^{(n)}\\
q^2\beta_2^{(n)}\\
q \beta_1^{(n)}\\
q^2(\beta_2^{(n)}-\beta_3^{(n)})\\
\end{matrix}\right).
\end{eqnarray}

\begin{eqnarray}
A_{nn} = \left( \begin{matrix}
1-\beta_0^{(n)}\widetilde{W}_1^{(nn,11)}-q^{2}\beta_2^{(n)}W_2^{(nn,1)}
& -\beta_0^{(n)}W_2^{(nn,1)}
& -\beta_1^{(n)}\alpha_3^{1n}-\beta_0^{(n)}\alpha_3^{0n}
&-C^{F}_{+}\beta_0^{(n)} \\
- C^{F}_{+}q^{2}(\beta_2^{(n)}-\beta_3^{(n)})+\beta_1^{(n)} \alpha_1^{1n} & & &\\
& & &\\
-q^2\beta_2^{(n)}\widetilde{W}_1^{(nn,11)}-q^{4}\beta_5^{(n)}W_2^{(nn,1)}
& 1-q^2\beta_2^{(n)}W_2^{(nn,1)}
&-q^2\beta_4^{(n)}\alpha_3^{1n}-q^2\beta_2^{(n)}\alpha_3^{0n}
&-q^2C^{F}_{+}\beta_2^{(n)} \\
- C^{F}_{+}q^{4}(\beta_5^{(n)}-\beta_8^{(n)})+q^2\beta_4^{(n)} \alpha_1^{1n}  & & & \\
& & &\\
-q\beta_1^{(n)}\widetilde{W}_1^{(nn,11)}-q^{3}\beta_4^{(n)}W_2^{(nn,1)}
&-q\beta_1^{(n)}W_2^{(nn,1)}
&1-q\beta_3^{(n)}\alpha_3^{1n}-q\beta_1^{(n)}\alpha_3^{0n}
&-qC^{F}_{+}\beta_1^{(n)} \\
 - C^{F}_{+}q^{3}(\beta_4^{(n)}-\beta_6^{(n)})+q\beta_3^{(n)} \alpha_1^{1n} & & & \\
& & &\\
-q^2(\beta_2^{(n)}-\beta_3^{(n)})\widetilde{W}_1^{(nn,11)}
&-q^2(\beta_2^{(n)}-\beta_3^{(n)})W_2^{(nn,1)}
& -q^2(\beta_4^{(n)}-\beta_6^{(n)})\alpha_3^{1n}
&1-C^{F}_{+}q^2(\beta_2^{(n)}-\beta_3^{(n)}) \\
 - C^{F}_{+}q^{4}(\beta_5^{(n)}-2\beta_8^{(n)}+\beta_7^{(n)})& & -q^2(\beta_2^{(n)}-\beta_3^{(n)})\alpha_3^{0n} & \\
 -q^{4}(\beta_5^{(n)}-\beta_8^{(n)})W_2^{(nn,1)} & & & \\
 +q^2(\beta_4^{(n)}-\beta_6^{(n)}) \alpha_1^{1n}   & & & \\
\end{matrix} \right)
\end{eqnarray}

\begin{eqnarray}
A_{np} = \left( \begin{matrix}
-\beta_0^{(n)}\widetilde{W}_1^{(np,11)}-q^{2}\beta_2^{(n)}W_2^{(np,1)}
& -\beta_0^{(n)}W_2^{(np,1)}
& -\beta_1^{(n)}\alpha_7^{1n}-\beta_0^{(n)}\alpha_7^{0n}
&-C^{F}_{-}\beta_0^{(n)} \\
- C^{F}_{-}q^{2}(\beta_2^{(n)}-\beta_3^{(n)})+\beta_1^{(n)} \alpha_5^{1n} & & &\\
& & &\\
-q^2\beta_2^{(n)}\widetilde{W}_1^{(np,11)}-q^{4}\beta_5^{(n)}W_2^{(np,1)}
& -q^2\beta_2^{(n)}W_2^{(np,1)}
&-q^2\beta_4^{(n)}\alpha_7^{1n}-q^2\beta_2^{(n)}\alpha_7^{0n}
&-q^2C^{F}_{-}\beta_2^{(n)} \\
- C^{F}_{+}q^{4}(\beta_5^{(n)}-\beta_8^{(n)})+q^2\beta_4^{(n)} \alpha_5^{1n}  & & & \\
& & &\\
-q\beta_1^{(n)}\widetilde{W}_1^{(np,11)}-q^{3}\beta_4^{(n)}W_2^{(np,1)}
&-q\beta_1^{(n)}W_2^{(np,1)}
&-q\beta_3^{(n)}\alpha_7^{1n}-q\beta_1^{(n)}\alpha_7^{0n}
&-qC^{F}_{-}\beta_1^{(n)} \\
 - C^{F}_{-}q^{3}(\beta_4^{(n)}-\beta_6^{(n)})+q\beta_3^{(n)} \alpha_5^{1n} & & & \\
& & &\\
-q^2(\beta_2^{(n)}-\beta_3^{(n)})\widetilde{W}_1^{(np,11)}
&-q^2(\beta_2^{(n)}-\beta_3^{(n)})W_2^{(np,1)}
& -q^2(\beta_4^{(n)}-\beta_6^{(n)})\alpha_7^{1n}
&-C^{F}_{-}q^2(\beta_2^{(n)}-\beta_3^{(n)}) \\
 - C^{F}_{-}q^{4}(\beta_5^{(n)}-2\beta_8^{(n)}+\beta_7^{(n)})& & -q^2(\beta_2^{(n)}-\beta_3^{(n)})\alpha_7^{0n} & \\
 -q^{4}(\beta_5^{(n)}-\beta_8^{(n)})W_2^{(np,1)} & & & \\
 +q^2(\beta_4^{(n)}-\beta_6^{(n)}) \alpha_5^{1n}   & & & \\
\end{matrix} \right)
\end{eqnarray}

\subsection{Channel S=1 M=0}

\begin{eqnarray}
X_{nn}= \left( \begin{matrix}
\langle G^{nn,10}_{RPA}\rangle\\
\langle k^{2} G^{nn,10}_{RPA}\rangle\\
\sqrt{\frac{4\pi}{3}}\langle kY_{10}G^{nn,10}_{RPA}\rangle\\
\frac{4\pi}{3}\langle k^2|Y_{10}|^2G^{nn,10}_{RPA}\rangle\\
\end{matrix}\right) \, , \,
X_{pn}= \left( \begin{matrix}
\langle G^{pn,10}_{RPA}\rangle\\
\langle k^{2}G^{pn,10}_{RPA}\rangle\\
\sqrt{\frac{4\pi}{3}}\langle kY_{10}G^{pn,10}_{RPA}\rangle\\
\frac{4\pi}{3}\langle k^2|Y_{10}|^2G^{pn,10}_{RPA}\rangle\\
\end{matrix}\right) \, , \,
B_n= \left(
\begin{matrix}
\beta_0^{(n)}\\
q^2\beta_2^{(n)}\\
q \beta_1^{(n)}\\
q^2\beta_3^{(n)}\\
\end{matrix}\right)
\end{eqnarray}

\begin{eqnarray}
A_{nn} = \left( \begin{matrix}
1-\beta_0^{(n)}\widetilde{W}_1^{(nn,10)}-q^{2}\beta_2^{(n)}W_2^{(nn,1)}
& -\beta_0^{(n)}W_2^{(nn,1)}
& -\beta_1^{(n)}\alpha_3^{1n}-\beta_0^{(n)}\alpha_3^{0n}
&-C^{F}_{+}\beta_0^{(n)} \\
- 2 C^{F}_{+}q^{2}\beta_3^{(n)}+\beta_1^{(n)} \alpha_1^{1n} & & &\\
& & &\\
-q^2\beta_2^{(n)}\widetilde{W}_1^{(nn,10)}-q^{4}\beta_5^{(n)}W_2^{(nn,1)}
& 1-q^2\beta_2^{(n)}W_2^{(nn,1)}
&-q^2\beta_4^{(n)}\alpha_3^{1n}-q^2\beta_2^{(n)}\alpha_3^{0n}
&-q^2C^{F}_{+}\beta_2^{(n)} \\
 - 2 C^{F}_{+}q^{4}\beta_8^{(n)}+q^2\beta_4^{(n)} \alpha_1^{1n} & & & \\
& & &\\
-q\beta_1^{(n)}\widetilde{W}_1^{(nn,10)}-q^{3}\beta_4^{(n)}W_2^{(nn,1)}
&-q\beta_1^{(n)}W_2^{(nn,1)}
&1-q\beta_3^{(n)}\alpha_3^{1n}-q\beta_1^{(n)}\alpha_3^{0n}
&-qC^{F}_{+}\beta_1^{(n)} \\
 - 2 C^{F}_{+}q^{3}\beta_6^{(n)}+q\beta_3^{(n)} \alpha_1^{1n} & & & \\
& & &\\
-q^2\beta_3^{(n)}\widetilde{W}_1^{(nn,10)}-q^{4}\beta_8^{(n)} W_2^{(nn,1)}
&-q^2\beta_3^{(n)}W_2^{(nn,1)}
&-q^2\beta_6^{(n)}\alpha_3^{1n}-q^2\beta_3^{(n)}\alpha_3^{0n}
&1/2-C^{F}_{+}q^2\beta_3^{(n)} \\
 - 2 C^{F}_{+}q^{4}\beta_7^{(n)}+q^2\beta_6^{(n)} \alpha_1^{1n}  & & & \\
\end{matrix} \right)
\end{eqnarray}

\begin{eqnarray}
A_{np} = \left( \begin{matrix}
 -\beta_0^{(n)}\widetilde{W}_1^{(np,10)}-q^{2}\beta_2^{(p)}W_2^{(np,1)}
& -\beta_0^{(n)}W_2^{(pn,1)}
& -\beta_1^{(n)}\alpha_7^{1n}-\beta_0^{(n)}\alpha_7^{0n}
& -C^{F}_{-}\beta_0^{(n)} \\
- 2 C^{F}_{-}q^{2}\beta_3^{(n)}+\beta_1^{(n)} \alpha_5^{1n} & & &\\
& & &\\
-q^2\beta_2^{(n)}\widetilde{W}_1^{(np,10)}-q^{4}\beta_5^{(n)}W_2^{(np,1)}
& -q^2\beta_2^{(n)}W_2^{(np,1)}
& -q^2\beta_4^{(n)}\alpha_7^{1n}-q^2\beta_2^{(n)}\alpha_7^{0n}
& -q^2C^{F}_{-}\beta_2^{(n)}\\
- 2C^{F}_{-}q^{4}\beta_8^{(n)}+q^2\beta_4^{(n)} \alpha_5^{1n}   & & & \\
& & &\\
-q\beta_1^{(n)}\widetilde{W}_1^{(np,10)}-q^{3}\beta_4^{(n)}W_2^{(np,1)}
& -q\beta_1^{(n)}W_2^{(np,1)}
& -q\beta_3^{(n)}\alpha_7^{1n}-q\beta_1^{(n)}\alpha_7^{0n}
& -qC^{F}_{-}\beta_1^{(n)} \\
- 2C^{F}_{-}q^{3}\beta_6^{(n)}+q^{}\beta_3^{(n)} \alpha_5^{1n} & & & \\
& & &\\
 -q^2\beta_3^{(n)}\widetilde{W}_1^{(np,10)}-q^{4}\beta_8^{(n)}W_2^{(np,1)}
& -q^2\beta_3^{(n)}W_2^{(np,1)}
& -q^2\beta_6^{(n)}\alpha_7^{1n}-q^2\beta_3^{(n)}\alpha_7^{0n}
& -C^{F}_{-}q^2\beta_3^{(n)} \\
 - 2C^{F}_{-}q^{4}\beta_7^{(n)}+q^2\beta_6^{(n)} \alpha_5^{1n}  & & & \\
\end{matrix} \right)
\end{eqnarray}

\section{Coefficients $\widetilde W_i^{(\tau,\tau',SM)}$ and $\alpha_i^{j \tau}$}
\label{app-wp}
The matrices of the previous appendix contain some short-hand notations which are defined in the following, channel by channel. We remind that coefficients $\widetilde W_i^{(\tau,\tau',SM)}$ are defined after partly solving the system of equations for the coupling between spin channels induced by the spin-orbit interaction. 

\subsection{Channel S=0 M=0}

\begin{eqnarray}\label{W1tildeS0}
{\widetilde{W}}_{1}^{(\tau\tau;00)}&=&W_{1}^{(\tau\tau,0)}-
4 q^4 \frac{C_{+}^{\nabla J}C_{-}^{\nabla J}(\beta_{2}^{-\tau}-\beta_{3}^{-\tau})x_{\tau-\tau}-[C_{+}^{\nabla J}]^{2}(\beta_{2}^{\tau}-\beta_{3}^{\tau}) (1+x_{-\tau-\tau})}{(1+x_{\tau\tau})({1+x_{-\tau-\tau}})-{x_{\tau-\tau}x_{-\tau\tau}}}\nonumber\\
&-&4q^4 \frac{C_{+}^{\nabla J}C_{-}^{\nabla J}{(\beta_{2}^{\tau}-\beta_{3}^{\tau})x_{-\tau\tau}}-[C_{-}^{\nabla J}]^{2}(\beta_{2}^{-\tau}-\beta_{3}^{-\tau}) ({1+x_{-\tau-\tau}})}
{(1+x_{\tau\tau})({1+x_{-\tau-\tau}})-{x_{\tau-\tau}x_{-\tau\tau}}} \, , \\
{\widetilde{W}}_{1}^{(\tau-\tau;00)}&=&W_{1}^{(\tau-\tau,0)}-
4q^4 \frac{[C_{+}^{\nabla J}]^{2}{(\beta_{2}^{-\tau}-\beta_{3}^{-\tau})x_{\tau-\tau}}-C_{-}^{\nabla J}C_{+}^{\nabla J}(\beta_{2}^{\tau}-\beta_{3}^{\tau}) ({1+x_{-\tau-\tau}})}
{(1+x_{\tau\tau})({1+x_{-\tau-\tau}})-{x_{\tau-\tau}x_{-\tau\tau}}}\nonumber\\
&-& 4q^4 \frac{[C_{-}^{\nabla J}]^{2}{(\beta_{2}^{\tau}-\beta_{3}^{\tau})x_{-\tau\tau}}-C_{-}^{\nabla J}C_{+}^{\nabla J}(\beta_{2}^{-\tau}-\beta_{3}^{-\tau})({1+x_{\tau\tau}} )}
{(1+x_{\tau\tau})({1+x_{-\tau-\tau}})-{x_{\tau-\tau}x_{-\tau\tau}}} \, ,
\end{eqnarray}
where
\begin{eqnarray}
x_{\tau \tau}=q^{2}(\beta_{2}^{\tau}-\beta_{3}^{\tau})(W_{2}^{(\tau \tau,1)}-C^{F}_{+}) \, , \\
x_{\tau -\tau}=q^{2}(\beta_{2}^{\tau}-\beta_{3}^{\tau})(W_{2}^{(\tau -\tau,1)}-C^{F}_{-}) \, .
\end{eqnarray}
It is worth stressing that the spin-orbit coupling constants, $C^{\nabla J}_{\pm}$, are associated to the power $q^4$.

\subsection{Channel S=1 M=1}

\begin{eqnarray}
\widetilde{W}_{1}^{(\tau \tau,11)}&=&W_{1}^{(\tau \tau,1)}-2q^{4}Z^{0}C^{\nabla J}_{+}\left[ z^{\tau -\tau,0}C^{\nabla J}_{-}(\beta_{2}^{-\tau}-\beta_{3}^{-\tau})-(1+z^{-\tau -\tau,0})C^{\nabla J}_{+}(\beta_{2}^{\tau}-\beta_{3}^{\tau})\right]\nonumber\\
&-&2q^{4}Z^{0}C^{\nabla J}_{-}\left[ z^{-\tau \tau,0}C^{\nabla J}_{+}(\beta_{2}^{\tau}-\beta_{3}^{\tau})-(1+z^{\tau \tau,0})C^{\nabla J}_{-}(\beta_{2}^{-\tau}-\beta_{3}^{-\tau})\right]\nonumber\\
&+&[C^{F}_{+}]^{2}q^{4}(\beta_{5}^{\tau}-\beta_{7}^{\tau})+[C^{F}_{-}]^{2}q^{4}(\beta_{5}^{-\tau}-\beta_{7}^{-\tau})\nonumber\\
&-&q^{2}\left( \frac{\nu_{\tau}}{k_{\tau}}-1\right)\left[ C^{F}_{+}(z^{\tau \tau,1}B^{-\tau \tau}_{+}+z^{\tau -\tau,1}B^{\tau -\tau}_{-})\right]-q^{2}\left( \frac{\nu_{-\tau}}{k_{-\tau}}-1\right)\left[C^{F}_{-}(z^{-\tau -\tau,1}B^{\tau -\tau}_{-}+z^{-\tau \tau,1}B^{-\tau \tau}_{+})\right]\,,\\
\widetilde{W}_{1}^{(\tau -\tau,11)}&=&W_{1}^{(\tau -\tau,1)}-2q^{4}Z^{0}C^{\nabla J}_{+}\left[ z^{\tau -\tau,0}C^{\nabla J}_{+}(\beta_{2}^{-\tau}-\beta_{3}^{-\tau})-(1+z^{-\tau -\tau,0})C^{\nabla J}_{-}(\beta_{2}^{\tau}-\beta_{3}^{\tau})\right]\nonumber\\
&-&2q^{4}Z^{0}C^{\nabla J}_{-}\left[ z^{-\tau\tau,0}C^{\nabla J}_{-}(\beta_{2}^{\tau}-\beta_{3}^{\tau})-(1+z^{\tau\tau,0})C^{\nabla J}_{+}(\beta_{2}^{-\tau}-\beta_{3}^{-\tau})\right]\nonumber\\
&+&C^{F}_{+}C^{F}_{-}q^{4}\left[ (\beta_{5}^{\tau}-\beta_{7}^{\tau})+(\beta_{5}^{-\tau}-\beta_{7}^{-\tau}) \right]\nonumber\\
&-&q^{2}\left( \frac{\nu_{\tau}}{k_{\tau}}-1\right)\left[ C^{F}_{+}(z^{\tau\tau,1}B^{-\tau\tau}_{-}+z^{\tau-\tau,1}B^{\tau-\tau}_{+})\right]-q^{2}\left( \frac{\nu_{-\tau}}{k_{-\tau}}-1\right)\left[C^{F}_{-}(z^{-\tau-\tau,1}B^{\tau-\tau}_{+}+z^{-\tau\tau,1}B^{-\tau\tau}_{-})\right]\,,
\end{eqnarray}

\noindent $\alpha_i ^{j \tau}$ is defined as the coefficient in front $\beta_j^{\tau}$ for the unknown parameter $X_i$ for the first four equations. They read
\begin{eqnarray*}
\alpha_1^{1n} & = & 2q^{2}\left[C^{F}_{+}B^{pn}_{+}+C^{F}_{-}B^{np}_{-} \right]\,,\\
\alpha_1^{1p} & = & 2q^{2}\left[C^{F}_{+}B^{np}_{-}+C^{F}_{-}B^{pn}_{+} \right]\,,\\
\alpha_3^{1n} & = & 2q(A^{pn}_{+}C^{F}_{+}+C^{F}_{-}A^{np}_{-})-2q W_{2}^{(nn,1)}\,, \\
\alpha_3^{1p} & = & 2q(A^{pn}_{+}C^{F}_{-}+C^{F}_{+}A^{np}_{-})-2q W_{2}^{(pn,1)}\,, \\
\alpha_3^{0n} & = & q \left(\frac{\nu_{n}}{k_{n}}-1\right) \left[C^{F}_{+}(z^{nn,1}A^{pn}_{+}+z^{np,1}A^{np}_{-})\right]+q \left(\frac{\nu_{p}}{k_{p}}-1\right)\left[C^{F}_{-}(z^{pp,1}A^{np}_{-}+z^{pn,1}A^{pn}_{+}) \right]\nonumber\\
&-& 2 q^{3}\left( [C^{F}_{+}]^{2}(\beta_{4}^{n}-\beta_{6}^{n})+[C^{F}_{-}]^{2}(\beta_{4}^{p}-\beta_{6}^{p})\right) \,,\\
\alpha_3^{0p} & = & q \left(\frac{\nu_{p}}{k_{p}}-1\right) \left[C^{F}_{+}(z^{pp,1}A^{np}_{-}+z^{pn,1}A^{pn}_{+})\right]+q \left(\frac{\nu_{n}}{k_{n}}-1\right)\left[C^{F}_{-}(z^{nn,1}A^{pn}_{+}+z^{np,1}A^{np}_{-}) \right]\nonumber\\
&-& 2 q^{3} C^{F}_{+}C_{-}^{F}\left[ (\beta_{4}^{n}-\beta_{6}^{n})+ (\beta_{4}^{p}-\beta_{6}^{p})\right] \,,\\
\alpha_5^{1n} & = & 2q^{2}\left[ C^{F}_{+}B^{pn}_{-}+C^{F}_{-}B^{np}_{+}\right]\,,\\
\alpha_5^{1p} & = & 2q^{2}\left[C^{F}_{+}B^{np}_{+}+C^{F}_{-}B^{pn}_{-} \right]\,,\\
\alpha_7^{1n} & = & 2q(A^{np}_{+}C^{F}_{-}+A^{pn}_{-}C^{F}_{+})-2q W_{2}^{(np,1)}\,,\\
\alpha_7^{1p} & = & 2q(A^{np}_{+}C^{F}_{+}+C^{F}_{-}A^{pn}_{-})-2q W_{2}^{(pp,1)} \,,\\
\alpha_7^{0n} & = & q \left(\frac{\nu_{n}}{k_{n}} -1\right)\left[ C^{F}_{+}(z^{nn,1}A^{pn}_{-}+z^{np,1}A^{np}_{+})\right]+q \left(\frac{\nu_{p}}{k_{p}} -1\right)\left[C^{F}_{-}(z^{pp,1}A^{np}_{+}+z^{pn,1}A^{pn}_{-})\right]\nonumber\\
&-&2C^{F}_{+}C^{F}_{-}q^{3}\left[ (\beta_{4}^{n}-\beta_{6}^{n})+(\beta_{4}^{p}-\beta_{6}^{p})\right]\,,\\
\alpha_7^{0p} & = & q \left(\frac{\nu_{p}}{k_{p}}-1\right) \left[C^{F}_{+}(z^{pp,1}A^{np}_{+}+z^{pn,1}A^{pn}_{-})\right]+q \left(\frac{\nu_{n}}{k_{n}} -1\right)\left[C^{F}_{-}(z^{nn,1}A^{pn}_{-}+z^{np,1}A^{np}_{+}) \right]\nonumber\\
&-& 2 q^{3}\left( [C^{F}_{+}]^{2}(\beta_{4}^{p}-\beta_{6}^{p})+[C^{F}_{-}]^{2}(\beta_{4}^{n}-\beta_{6}^{n})\right)\,.
\end{eqnarray*}

In these expressions
\begin{eqnarray}\label{zdefinition}
z^{\tau \tau',S}&=&q^{2}(\beta_{2}^{\tau}-\beta_{3}^{\tau})W_{2}^{\tau \tau',S},\\
Z^{S}&=&\left[(1+z^{nn,S})(1+z^{pp,S})-z^{pn,S}z^{np,S}\right]^{-1}.
\end{eqnarray}
\begin{eqnarray}
A^{\tau \tau'}_{\pm}&=&q^{2}Z^{1}\left[ (1+z^{\tau \tau,1})C^{F}_{\pm}(\beta_{2}^{\tau'}-\beta_{3}^{\tau'})-C^{F}_{\mp} z^{\tau' \tau,1}(\beta_{2}^{\tau}-\beta_{3}^{\tau})\right]\,,\\
B^{\tau \tau'}_{\pm}&=&q^{2}Z^{1}\left[ (1+z^{\tau \tau,1})C^{F}_{\pm}(\beta_{4}^{\tau'}-\beta_{6}^{\tau'})-C^{F}_{\mp} z^{\tau' \tau,1}(\beta_{4}^{\tau}-\beta_{6}^{\tau})\right]\,.
\end{eqnarray}

The other four equations for the system can be deduced from the first four ones by changing isospin indices 
like $n \rightarrow p$ together with the following substitutions:

\begin{eqnarray*}
\alpha_{1}^{1n} \longleftrightarrow \alpha_{5}^{1p}  \,, \\
\alpha_{3}^{1n} \longleftrightarrow \alpha_{7}^{1p} \,, \\
\alpha_{3}^{0n} \longleftrightarrow \alpha_{7}^{0p}  \,, \\
\alpha_{5}^{1n} \longleftrightarrow \alpha_{1}^{1p}\, , \\
\alpha_{7}^{1n} \longleftrightarrow \alpha_{3}^{1p}\, , \\
\alpha_{7}^{0n} \longleftrightarrow \alpha_{3}^{0p}  \,.
\end{eqnarray*}

\subsection{Channel S=1 M=0}

\begin{eqnarray}
\widetilde{W}^{(\tau\tau,10)}_{1}&=&W_{1}^{(\tau\tau,1)}-q^{2}C^{F}_{+}\left( \frac{\nu_{\tau}}{k_{\tau}}-1\right)\left[\bar{B}^{-\tau\tau}_{+}(z^{\tau\tau,1}+3s^{\tau}_{+})+\bar{B}^{\tau-\tau}_{-}(z^{\tau-\tau,1}+3s^{\tau}_{-}) \right]\\
&-&q^{2}C^{F}_{-}\left( \frac{\nu_{-\tau}}{k_{-\tau}}-1\right)\left[\bar{B}^{-\tau\tau}_{+}(z^{-\tau\tau,1}+3s^{-\tau}_{-})+\bar{B}^{\tau-\tau}_{-}(z^{-\tau-\tau,1}+3s^{-\tau}_{+}) \right]\\
&+&4q^{4}\left[(C^{F}_{+})^{2}(\beta_{8}^{\tau}-\beta_{7}^{\tau})+(C^{F}_{-})^{2}(\beta_{8}^{-\tau}-\beta_{7}^{-\tau}) \right]\, ,\nonumber\\
\widetilde{W}^{(\tau-\tau,10)}_{1}&=&W_{1}^{(\tau-\tau,1)}-q^{2}C^{F}_{+}\left( \frac{\nu_{\tau}}{k_{\tau}}-1\right)\left[\bar{B}^{\tau-\tau}_{+}(z^{\tau-\tau,1}+3s^{\tau}_{-})+\bar{B}^{-\tau\tau}_{-}(z^{\tau\tau,1}+3s^{\tau}_{+}) \right]\\
&-&q^{2}C^{F}_{-}\left( \frac{\nu_{-\tau}}{k_{-\tau}}-1\right)\left[\bar{B}^{\tau-\tau}_{+}(z^{-\tau-\tau,1}+3s^{-\tau}_{+})+\bar{B}^{-\tau\tau}_{-}(z^{-\tau\tau,1}+3s^{-\tau}_{-}) \right]\\
&+&4q^{4}C^{F}_{+}C^{F}_{-}\left[(\beta_{8}^{\tau}-\beta_{7}^{\tau})+(\beta_{8}^{-\tau}-\beta_{7}^{-\tau}) \right]\, .\nonumber
\end{eqnarray}

\noindent The coefficients $\alpha_i ^{l,n(p)}$ are different for this channel. They read explicitly

\begin{eqnarray}
\alpha^{1n}_{1}&=&2q^{2}\left[C^{F}_{+}\bar{B}^{pn}_{+}+C^{F}_{-}\bar{B}^{np}_{-} \right]\, ,\\
\alpha^{1p}_{1}&=&2q^{2}\left[ C^{F}_{+}\bar{B}^{np}_{-}+C^{F}_{-}\bar{B}^{pn}_{+}\right]\, ,\\
\alpha^{1n}_{3}&=&2q\left[-W_{2}^{nn,1}-2C^{F}_{+}+2\bar{A}^{pn}_{+}C^{F}_{+}+2\bar{A}^{np}_{-}C^{F}_{-}\right]\, ,\\
\alpha^{1p}_{3}&=&2q\left[ -W_{2}^{pn,1}-2C^{F}_{-}+2C^{F}_{+}\bar{A}^{np}_{-}+2C^{F}_{-}\bar{A}^{pn}_{+}\right]\, ,\\
\alpha^{0n}_{3}&=&2q\left(\frac{\nu_{n}}{k_{n}}-1\right)C^{F}_{+}\left[ \bar{A}^{pn}_{+}(z^{nn,1}+3s^{n}_{+})+\bar{A}^{np}_{-}(z^{np,1}+3s^{n}_{-})-s^{n}_{+}\right]\nonumber\\
&+& 2q\left(\frac{\nu_{p}}{k_{p}}-1\right)C^{F}_{-}\left[ \bar{A}^{np}_{-}(z^{pp,1}+3s^{p}_{+})+\bar{A}^{pn}_{+}(z^{pn,1}+3s^{p}_{-})-s^{p}_{-}\right]\, ,\\
\alpha^{0p}_{3}&=& 2q\left( \frac{\nu_{p}}{k_{p}}-1\right)C^{F}_{+}\left[\bar{A}^{np}_{-}(z^{pp,1}+3s^{p}_{+})+\bar{A}^{pn}_{+}(z^{pn,1}+3s^{p}_{-}) -s^{p}_{-}\right]\nonumber\\
&+&2q\left( \frac{\nu_{n}}{k_{n}}-1\right)C^{F}_{-}\left[\bar{A}^{pn}_{+}(z^{nn,1}+3s^{n}_{+})+\bar{A}^{np}_{-}(z^{np,1}+3s^{n}_{-}) -s^{n}_{+}\right]\, ,\\
\alpha^{1n}_{5}&=&2q^{2}\left[ C^{F}_{+}\bar{B}^{pn}_{-}+C^{F}_{-}\bar{B}^{np}_{+}\right]\, ,\\
\alpha^{1p}_{5}&=&2q^{2}\left[ C^{F}_{+}\bar{B}^{np}_{+}+C^{F}_{-}\bar{B}^{pn}_{-} \right]\, ,\\
\alpha^{1n}_{7}&=&2q\left[ -W_{2}^{np,1}-2C^{F}_{-}+2C^{F}_{+}\bar{A}^{pn}_{-}+2C^{F}_{-}\bar{A}^{np}_{+}\right] \, ,\\
\alpha^{1p}_{7}&=&2q\left[-W_{2}^{pp,1}-2C^{F}_{+}+2\bar{A}^{np}_{+}C^{F}_{+}+2\bar{A}^{pn}_{-}C^{F}_{-}\right] \, ,\\
\alpha^{0n}_{7}&=&2q\left( \frac{\nu_{n}}{k_{n}}-1\right)C^{F}_{+}\left[\bar{A}^{pn}_{-}(z^{nn,1}+3s^{n}_{+})+\bar{A}^{np}_{+}(z^{np,1}+3s^{n}_{-}) -s^{n}_{-}\right]\nonumber\\
&+&2q\left( \frac{\nu_{p}}{k_{p}}-1\right)C^{F}_{-}\left[\bar{A}^{np}_{+}(z^{pp,1}+3s^{p}_{+})+\bar{A}^{pn}_{-}(z^{pn,1}+3s^{p}_{-}) -s^{p}_{+}\right]\, ,\\
\alpha^{0p}_{7}&=&2q\left(\frac{\nu_{p}}{k_{p}}-1\right)C^{F}_{+}\left[ \bar{A}^{np}_{+}(z^{pp,1}+3s^{p}_{+})+\bar{A}^{pn}_{-}(z^{pn,1}+3s^{p}_{-})-s^{p}_{+}\right]\nonumber\\
&+&2q\left(\frac{\nu_{n}}{k_{n}}-1\right)C^{F}_{-}\left[ \bar{A}^{pn}_{-}(z^{nn,1}+3s^{n}_{+})+\bar{A}^{np}_{+}(z^{np,1}+3s^{n}_{-})-s^{n}_{-}\right]\, .
\end{eqnarray}

\noindent The transformation rules to get the last four equations are identical to the ones of channel $S=1, M=1$.

\begin{eqnarray}\label{esse}
s_{\pm}^{\tau}&=&C^{F}_{\pm}q^{2}(\beta_{2}^{\tau}-\beta_{3}^{\tau})\, ,\\
Z'&=&\left[(1+z^{pp,1}+3s^{p}_{+})(1+z^{nn,1}+3s^{n}_{+})-(z^{np,1}+3s^{n}_{-})(z^{pn,1}+3s^{p}_{-})\right]^{-1}\, ,
\end{eqnarray}

\begin{eqnarray}
\bar{A}^{\tau\tau'}_{\pm}&=&Z'\left[(1+z^{\tau\tau,1}+3s^{\tau}_{+})s^{\tau'}_{\pm}-(z^{\tau'\tau,1}+3s^{\tau'}_{-})s^{\tau}_{\mp} \right]\, ,\\
\bar{B}^{\tau\tau'}_{+}&=&Z'\left[\left(\frac{\nu_{\tau'}}{k_{\tau'}}-1\right)(1+z^{\tau\tau,1}+3s^{\tau}_{+})s^{\tau'}_{+}-\left(\frac{\nu_{\tau}}{k_{\tau}}-1\right)(z^{\tau'\tau,1}+3s^{\tau'}_{-})s^{\tau}_{-} \right]\, .
\end{eqnarray}

\end{appendix}


\end{document}